\documentclass[journal=jacs,manuscript=article]{achemso}
\usepackage[version=3]{mhchem}
\usepackage{xr}
\usepackage{outlines}
\setkeys{acs}{layout = traditional}
\setkeys{acs}{usetitle = true}
\usepackage{graphicx,dcolumn,bm,amsmath,amssymb}
\usepackage[usenames, dvipsnames]{color}
\usepackage[labelfont=bf]{caption}
\usepackage{pslatex}
\usepackage{tabu}
\usepackage{makecell}
\usepackage{textcomp}
\usepackage{amsmath}
\newcommand*\mycommand[1]{\texttt{\emph{A1}}}
\newcommand{\fstar}{$f^{*}$\ }
\newcommand{\hidden}[1]{}
\graphicspath{ {./figures/} }

\makeatletter
\newcommand*{\addFileDependency}[1]{
  \typeout{(#1)}
  \@addtofilelist{#1}
  \IfFileExists{#1}{}{\typeout{No file #1.}}
}
\makeatother

\newcommand*{\myexternaldocument}[1]{%
    \externaldocument{#1}%
    \addFileDependency{#1.tex}%
    \addFileDependency{#1.aux}%
}

\myexternaldocument{supp}

\title{Selectivity in yttrium manganese oxide synthesis via local chemical potentials in hyperdimensional phase space}

\author{Paul K. Todd}
\affiliation{Department of Chemistry, Colorado State University,  Fort Collins, Colorado 80523-1872, USA}
\altaffiliation{These authors contributed equally to this work.}
\author{Matthew J. McDermott}
\affiliation{Materials Sciences Division, Lawrence Berkeley National Laboratory, 1 Cyclotron Road, Berkeley, CA 94720, USA}
\alsoaffiliation{Department of Materials Science and
Engineering, University of California, Berkeley,
CA 94720, USA}
\altaffiliation{These authors contributed equally to this work.}
\author{Christopher L. Rom}
\affiliation{Department of Chemistry, Colorado State University, Fort Collins, Colorado 80523-1872, USA}
\author{Adam A. Corrao}
\affiliation{Department of Chemistry, Stony Brook University, Stony Brook, NY 11794-3400, USA}
\author{Jonathan J. Denney}
\affiliation{Department of Chemistry, Stony Brook University, Stony Brook, NY 11794-3400, USA}
\author{Shyam S. Dwaraknath}
\affiliation{Materials Sciences Division, Lawrence Berkeley National Laboratory, 1 Cyclotron Road, Berkeley, CA 94720, USA}
\author{Peter G. Khalifah}
\affiliation{Department of Chemistry, Stony Brook University, Stony Brook, NY 11794-3400, USA}
\alsoaffiliation{Department of Chemistry, Brookhaven National Laboratory, Upton, NY 11973, USA}
\author{Kristin A. Persson}
\affiliation{Molecular Foundry, Lawrence Berkeley National Laboratory, 1 Cyclotron Road, Berkeley, CA 94720, USA}
\alsoaffiliation{Department of Materials Science and Engineering, University of California, Berkeley, CA 94720, USA}
\author{James R. Neilson}
\affiliation{Department of Chemistry, Colorado State University, Fort Collins, Colorado 80523-1872, USA}
\email{james.neilson@colostate.edu}

\begin{document}
\newpage
\begin{abstract}

In sharp contrast to molecular synthesis, materials synthesis is generally presumed to lack selectivity. The few known methods of designing selectivity in solid-state reactions have limited scope, such as topotactic reactions or strain stabilization. This contribution describes a general approach for searching large chemical spaces to identify selective reactions. This novel approach explains the ability of a nominally ``innocent'' \ce{Na2CO3} precursor to enable the metathesis synthesis of single-phase \ce{Y2Mn2O7} -- an outcome that was previously only accomplished at extreme pressures and which cannot be achieved with closely related precursors of \ce{Li2CO3} and \ce{K2CO3} under identical conditions.  By calculating the required change in chemical potential across all possible reactant-product interfaces in an expanded chemical space including Y, Mn, O, alkali metals, and halogens, using thermodynamic parameters obtained from density functional theory calculations, we identify reactions that minimize the thermodynamic competition from intermediates. In this manner, only the Na-based intermediates minimize the distance in the hyperdimensional chemical potential space to \ce{Y2Mn2O7}, thus providing selective access to a phase which was previously thought to be metastable. Experimental evidence validating this mechanism for pathway-dependent selectivity is provided by intermediates identified from \textit{in situ} synchrotron-based crystallographic analysis.  This approach of calculating chemical potential distances in hyperdimensional compositional spaces provides a general method for designing selective solid-state syntheses that will be useful for gaining access to metastable phases and for identifying reaction pathways that can reduce the synthesis temperature, and cost, of technological materials.
\end{abstract}
\newpage

\section{Introduction}

Molecular reactions use catalysts to increase the rate of reaction and achieve selectivity by modifying specific barriers in the reaction pathway. However, this approach cannot be readily translated to the synthesis of non-molecular compounds, which are found in a wide range of industrial applications ranging from batteries in electronic devices to cement in concrete. To achieve selective control over solid-state reactions, a necessary first step is ``turning down the heat'' enough to prevent non-selective phase interconversion \cite{Stein1993}. An extreme example of this strategy is a topotactic reaction involving ion exchange in which only a single type of atom is mobile (e.g., cations \cite{Murphy1977} or anions \cite{Tsujimoto2007,Yajima:2015db}), which has enabled lithium-ion battery technology \cite{Armand2008}. However, such reactions require a relatively inert framework that facilitates the mobility of these ions. Another route to selective synthesis is epitaxial strain stabilization on a single-crystal or polycrystalline substrate \cite{Havelia2013}. While algorithms to search for suitable substrates exist \cite{Ding2016}, the epitaxial stabilization requires the synthesis of a chemically-compatible lattice-matched substrate, which yields only a limited amount of material that is necessarily attached to the underlying substrate.

To address the challenge of designing selective reactions in solid-state chemistry, one must have an understanding of how the reaction proceeds. Pairwise reactions at interfaces between solids dominate selection of a reaction pathway \cite{Miura2020b, Bianchini2020}. With the inclusion of additional elements, as in metathesis reactions \cite{Bonneau1991,wiley1992rapid}, one can avoid highly stable and thus unreactive intermediates, as demonstrated in the formation of \ce{MgCr2S4} while avoiding the inert reactant \ce{Cr2S3} \cite{Miura2020a, Wustrow2018}. In the preparation of complex oxides, ``assisted'' metathesis reactions reach quantitative yields at lower temperatures than conventional ceramic reactions,\cite{Seshadri2012} and they can even exhibit product selectivity depending on which alkali (A = Li, Na, K) carbonate precursor is used. This is seen for the reactions of \ce{3A2CO3 + 2YCl3 + Mn2O3}, where in the case of \ce{Li2CO3} the reaction selectively forms orthorhombic \ce{YMnO3}, \ce{Na2CO3} yields \ce{Y2Mn2O7}, and \ce{K2CO3} does not yield selectivity \cite{Todd2019}.

A central challenge of this contribution is to understand the origin of the different product outcomes that result from changing an ancillary precursor that is not directly involved in any elementary reaction producing a yttrium manganese oxide. In conventional stoichiometric reactions using binary yttrium oxide and manganese oxide precursors, hexagonal \ce{YMnO3}, denoted h-\ce{YMnO3}, forms at temperatures greater than 950 \textcelsius{} \cite{Kamata1979,golikov}. Previous reports have described the synthesis of \ce{Y2Mn2O7} under high temperatures and highly oxidizing conditions (1100 \textcelsius{} and 4 GPa with \ce{KIO4 -> KIO3 + 1/2 O2})\cite{Fujinaka1979} or hydrothermally under oxidizing conditions (NaOH, \ce{NaClO3} in \ce{H2O} at 500 \textcelsius{} and 3 GPa)\cite{Subramanian1988}, leading to its previous characterization as a metastable phase \cite{Gardner2010a}. Interestingly, thermogravimetric analysis shows the \ce{Y2Mn2O7} pyrochlore to be stable up to 800 \textcelsius{}, above which it decomposes to the perovskite \ce{YMnO3}.\cite{Subramanian1988} Therefore, if \ce{Y2Mn2O7} is stable at the lower temperatures, its stability cannot solely explain why only the sodium-based assisted metathesis precursors are selective for its formation given the low reaction temperatures for all of the alkali metathesis reactions. 

Here, using temperature-dependent synchrotron powder X-ray diffraction (SPXRD) experiments to study the \ce{Y2Mn2O7}-forming assisted metathesis reaction with a sodium precursor, \ce{Mn2O3 + 2\text{~}YCl3 + 3\text{~}Na2CO3 + 1/2\text{~}O2 -> Y2Mn2O7 + 6\text{~}NaCl + 3\text{~}CO2}, we identify a specific reaction pathway dependent on the formation of \ce{Y2O3} and its reaction with \ce{Na_{$x$}MnO2} intermediates. Density functional theory (DFT) derived thermochemistry reveals that \ce{Y2Mn2O7} is thermodynamically stable below $T\approx$ 1100 \textcelsius{} and that sodium-based intermediates provide selective formation of this phase by circumventing the formation of other Y-Mn-O intermediates (e.g., \ce{YMnO3} and \ce{YMn2O5}).  This sodium based reaction differs from the equivalent reactions with \ce{Li2CO3} and \ce{K2CO3}, which we previously showed instead yield \ce{YMnO3}, with a reaction temperature as low as 500 \textcelsius{}  \cite{Todd2019a,Todd2020}. We attribute the selective formation of \ce{Y2Mn2O7} to the distance of the Na-based intermediates in chemical potential space, which is shown visually on the chemical potential diagram and is numerically calculated between the stability regions of the reactants and products in a reaction. These methods, applied to easily retrievable materials data, now allow one to search for selective reactions over large, hyperdimensional compositional spaces.

\section{Methods}
\subsection{Experimental methods}
All reagents were prepared, stored, and weighed in an argon-filled glovebox with \ce{O2} and \ce{H2O} levels $\leq$ 0.1ppm. Manganese(III) oxide (Sigma Aldrich 99\%) was purified by heating \ce{Mn2O3} in an alumina boat at 1 \textcelsius{}/min to 700 \textcelsius{} for 16 h in air and quenched into the glovebox; purity was verified by powder X-ray diffraction (PXRD). Manganese (IV) oxide (Sigma Aldrich $\geq$ 99\%), \ce{YCl3} (Alfa Aesar 99.9\%), sodium carbonate (ACROS Organics 99.5\%), and manganese (II) carbonate (Sigma Aldrich $\geq$ 99.99\%) were purchased and stored under argon. All gases (\ce{O2}, \ce{Ar}, He) were purchased through Airgas at the Ultra High Purity grade (99.999\%). \ce{NaMnO2} ($C2/m$) was prepared by mixing manganese (III) oxide and sodium carbonate in a 1:1 molar ratio, grinding for 15 minutes in an agate mortar and pestle, and pelleting using a $1/4$ in die under $\sim$1 tn of force. The pellet was placed upon a sacrificial layer of powder in an alumina crucible and heated in a muffle furnace at 1 \textcelsius{}/min to 700 \textcelsius{} for 10 h. The reaction was subsequently quenched into the antechamber of the glovebox and stored under argon. Similarly, \ce{Na_{0.7}MnO2} ($Cmcm$) was prepared by mixing manganese (II) carbonate and sodium carbonate in a 0.70:1 Na:Mn  ratio following the same pellet preparation as \ce{NaMnO2}. The reaction was performed by heating at 5 \textcelsius{}/min to 1050 \textcelsius{} for 15 h and then quenching into the antechamber of a glovebox and stored under argon. \ce{Y2O3} was purified of hydroxide in an alumina boat at 1 \textcelsius{}/min to 900 \textcelsius{} for 4 h in air, cooled to 200 \textcelsius{} at 1 \textcelsius{}/min for 12 h, and quenching into the glovebox. \ce{YOCl} was prepared by heating \ce{YCl3*6H2O} in an alumina boat to 350 \textcelsius{} at 10 \textcelsius{}/min for 4 h in air. The \ce{YOCl} product formed as the $P4/nmn$ \ce{PbClF} structure-type. All prepared reactants were confirmed using laboratory PXRD. Preparations for $ex$ $situ$ assisted metathesis reactions have been described in detail previously where reaction yields match the expected  \ce{Y2Mn2O7}:6NaCl by PXRD (14.2(3) mol\% \ce{Y2Mn2O7}).\cite{Todd2019} Laboratory PXRD data were collected on a Bruker D8 Discover diffractometer using Cu K$\alpha$ radiation and a Lynxeye XE-T position-sensitive detector. 

For temperature-dependent \textit{in situ} assisted metathesis reactions that produce carbon dioxide as a by-product, open-ended quartz capillaries (1.1 mm OD) were packed in a glove-bag under argon using glass wool as a plug. Synchrotron X-ray diffraction experiments were performed at beamline 17-BM-B ($\lambda$ = 0.2415 \AA) at the Advanced Photon Source (APS) at Argonne National Laboratory using a Perkin Elmer plate detector at a distance of 700 mm. All capillaries were loaded into a flow-cell apparatus equipped with resistive heating elements and heated at 10 \textcelsius{}/min.\cite{Chupas2008} Gas flow (\ce{O2}, He) was controlled through mass flow controllers at a rate of 0.2 cc/min. Assisted-metathesis reactions were heated to a maximum temperature of 850 \textcelsius{} while the sample continuously rocked at $\pm$ 5$^\circ$ around the axis of the capillary. Diffraction patterns were collected every two seconds and summed every 20 seconds for powder averaging. Plate detector images were integrated using GSAS-II and calibrated using a \ce{LaB6} standard. 

All Rietveld refinements were performed using TOPAS v6. Due to the number and positional overlap of intermediates during the sequential refinements, thermal displacement parameters were fixed at 5\AA$^2$ and the full-width-half-max using a Lorentzian polynomial was fixed at 178 nm to better account for changes in peak intensity during the reaction. In order to compare the relative fractions of phases determined from Rietveld calculations\cite{Jiang2016}, a weighted scale factor (WSF) is defined as: $Q_{p} = S_{p} \cdot V_{p} \cdot M_{p}$ where $Q_{p}$ = weighted scale factor of phase p, $S_{p}$ = Scale factor calculated from Rietveld for phase p, $V_{p}$ is the volume of the unit cell for phase p, and $M_{p}$ is the atomic mass of the unit cell for phase p. It should be noted that we omit the Brindley coefficient for microabsorption correction in our calculation of weighted scale factor due to the unreliable refinement of particle sizes for individual phases. Amorphous material and product lost as vapor are not accounted for in the sequential refinement. We reference all phases by their nominal stoichiometric formula; however, the actual chemical formula may be distinct from the written formula as XRD data alone cannot typically resolve non-stoichiometric compounds.

It has been recently demonstrated by some of us that \fstar diagrams are a powerful tool for understanding defects in chemical systems with three different crystallographic sites, allowing a diffraction analog of a ternary phase diagram to be calculated in which each axis represents the relative scattering power of each crystallographic site (at 2$\theta = 0$) rather than the chemical composition.\cite{Yin2018, Yin2020} The \fstar diagram method is applied here with a slight modification to understand the evolution of defects within the \ce{Na_{x}MnO2} intermediates. The $C2/m$ P3 structure has three distinct crystallographic sites (one for each element), whereas the $Cmcm$ Birnessite P2 phase has four distinct crystallographic sites (2 Na, 1 Mn, 1 O) rather than three. Therefore, the scattering powers of the two Na sites were summed together in the construction of the \fstar diagram.

\subsection{Computational methods}
\subsubsection{Calculating thermodynamic free energies}
DFT-based atomic structures and formation enthalpies for material phases in the Y-Mn-O-Na-Li-K chemical systems were acquired from the Materials Project (MP) database, version 2021\.05\.13 \cite{Jain2013} Gibbs free energies of formation, $\Delta_f G^0(T)$, of solid DFT compounds were estimated using the machine-learned Gibbs free energy descriptor approach implemented by Bartel, et al. \cite{Bartel2018} and applied to MP data. The pymatgen package was used to perform convex hull phase stability analysis. \cite{Ong2010}

To model the energetics of reactions occurring in flowing \ce{O2} gas, the grand potential energy, $\Phi$, was used as the relevant thermodynamic potential:

\begin{equation}
    \Phi = G(P,T) - \mu_{\text{O}} N_{\text{O}}
\end{equation}

\noindent where $\Phi$ is normalized to the number of non-oxygen atoms. Since in place of $G(P,T)$ we used standard Gibbs free energies of formation, $\Delta G_f^0$(T=650\textcelsius{}), the chemical potential of oxygen was assigned a value of zero, i.e., $\mu_{\text{O}} = \mu_{\text{O}}^0$ ($p=0.1$ MPa, $T=650$ \textcelsius{}).  While this did not result in a shift in the total grand potential energy of any particular phase, it did change the number of atoms used for normalization and hence affected the magnitude of the normalized reaction energies, $\Delta \Phi_\mathrm{rxn}$. 

\subsubsection{Construction of chemical potential diagrams}
Chemical potential diagrams, as well as their more traditional two-dimensional versions (predominance diagrams), were constructed using the methodology described by Yokokawa. \cite{Yokokawa1999} We used an algorithmic approach inspired by a similar method for construction of Pourbaix diagrams,\cite{Montoya2019} which is briefly summarized here.

For a pure substance with $N$ atoms per chemical formula unit, consisting of $n$ unique elements indexed $i$ with concentrations of $x_i$ , its hyperplane in $n$-dimensional chemical space is defined via the equation:

\begin{equation}
    \sum_{i=1}^{n}{x_i(\mu_i - \mu^0_i)} = \frac{1}{N}\Delta G_f^0(T)
\end{equation}

\noindent where the standard Gibbs free energy of formation, $\Delta G_f^0(T)$, has been normalized to an energy-per-atom basis by dividing by the total number of atoms per formula unit, $N$. Note that all reference chemical potentials $\mu^0_i$ are set to zero when working with elements in their standard states.

Following the construction of hyperplanes for all phases in the chemical system, the stability domains are acquired by taking the convex hull of all points belonging to that phase in the set of intersections of the lowest energy hyperplanes, as calculated with the HalfspaceIntersection code implemented in SciPy. \cite{Virtanen2020} In this construction, each stability region is a convex $n-1$ dimensional polytope in $n$-dimensional chemical potential space. Since many possible chemical reactions involve more than $n=3$ elements, we must take additional steps to visualize the hyperdimensional ($n>3$) phase boundaries relevant to the reactions in this work. This can be accomplished by 1) using clever transformations of the axes, 2) setting one or more chemical potentials to fixed values, or 3) taking a slice of the relevant chemical potential polytopes in lower dimensions. Here we choose the third option, which uniquely allows for comparison across the full Y-Mn-O-Na-Li-K-Cl assisted metathesis system. Phases that do not directly appear on the generalized chemical potential diagram in three dimensions, e.g., \ce{AMnO2} in the Y-Mn-O subspace, are thus computed first in higher dimension chemical potential space (A-Y-Mn-O) and then sliced to plot in lower-dimensional subspace. In the case of \ce{AMnO2}, these regions appear as three-dimensional polyhedra due to the added degree of freedom in the chemical potential of the alkali element, $\mu_{\text{A}}$. These polyhedra also at least partially contact the stability areas of Y-Mn-O compounds, indicating where phases are adjacent in higher-dimensional space.

The ``chemical potential distance'', $\Delta \mu_{\mathrm{min}}(P_a, P_b)$, between any two phases $P_a$ and $P_b$ was calculated geometrically by finding the minimum Euclidean distance between the domains (i.e., the convex stability polytopes) of each phase on the chemical potential diagram. This was accomplished practically by computing the minimum Euclidean distance between all pairs of vertices ($\boldsymbol{\mu}_{a_i}$, $\boldsymbol{\mu}_{b_j}$) defining each of the two convex polytopes of phases $P_a$ and $P_b$:

\begin{equation}
\Delta \mu_{\mathrm{min}}(P_a, P_b) = \min_{i, j}\left\{\lVert \boldsymbol{\mu}_{a_i} - \boldsymbol{\mu}_{b_j} \rVert\right\}
\end{equation}

\noindent While the minimum distance between vertices is not necessarily equivalent to the minimum distance between any facet of the polytopes, the implementation of the more accurate distance calculation is nontrivial. We thus found the vertex based k-d tree method, as implemented in SciPy, to be an appropriate trade-off between computational complexity and accuracy.

\subsubsection{Enumerating reactions with \ce{Y2Mn2O7} as a product}

Chemical reactions with \ce{Y2Mn2O7} as a product were enumerated via a network-based approach described in a previous work \cite{McDermott2021} and implemented in the reaction-network Python package. \cite{zenodo} To summarize, the enumeration approach is a combinatorial brute force approach whereby every combination of phases up to a size of $n$ is considered both as a possible set of reactants or products. The set of all possible reactions is calculated by taking the Cartesian product between the set of reactant phase combinations and product phase combinations and filtering by whether it is possible to write a stoichiometrically balanced reaction equation between the reactants and products. In this work, reactions up to a size $n=2$ are enumerated from the set of 2,878 phases predicted to be stable at $T=650$\textcelsius{} within the alkali metal (Li, Na, K, Rb, Cs), halogen (F, Cl, Br, I), carbon (C), and target (Y, Mn, O) chemical system. To account for the system being open to oxygen gas, we also consider reaction equations with oxygen as an optional reactant or product beyond the $n=2$ limitation.

The enumerated reactions were compared against each other by using both the normalized reaction free energy and a ``total'' chemical potential distance for the reaction as ``cost'' variables. For the purposes of ranking reactions, the total cost of a reaction, $C$, was calculated using the softplus function with an equally weighted mean between the two cost variables:

\begin{equation}
C = \ln\left(1 + \frac{273}{T}\exp\left(\frac{\Delta \Phi_{\textrm{rxn}}+ \sum \Delta \mu_{\textrm{min}}}{2}\right)\right)
\label{softplus}
\end{equation}

\noindent where $T$ corresponds to a temperature of $T=923$ K, the reaction energy, $\Delta \Phi_\mathrm{rxn}$, corresponds to the normalized change in grand potential energy, and $\sum \Delta \mu_{\textrm{min}}$ represents the total chemical potential distance for a reaction, which is calculated as the sum of the (minimum) chemical potential distances between all possible pairwise interfaces in the reaction, excluding the interface(s) between the reactants. For example, in the hypothetical reaction A+B$ \xrightarrow{}$ C+D, the total chemical potential distance is calculated by taking the summation of $\Delta \mu_{\textrm{min}}$(A, C), $\Delta \mu_{\textrm{min}}$(A, D), $\Delta \mu_{\textrm{min}}$(B, C), $\Delta \mu_{\textrm{min}}$(B, D), and $\Delta \mu_{\textrm{min}}$(C, D). 

The full set of enumerated chemical reactions, as well as all experimental raw data, processed data, data processing scripts, and figure plotting scripts, are available at \url{https://github.com/GENESIS-EFRC/y2mn2o7-selectivity}. The entire reaction enumeration and ranking approach has also been implemented within the aforementioned reaction-network Python package \cite{zenodo} and can be applied in a similar fashion to recommend precursors for synthesizing other target materials. The applicability of this approach beyond the Y-Mn-O system is the subject of future work(s).

\section{Results and Discussion} 

\subsection{Reaction pathway from phase evolution}

\begin{figure}[ht!]
\begin{center}
\includegraphics[width=5.0in]{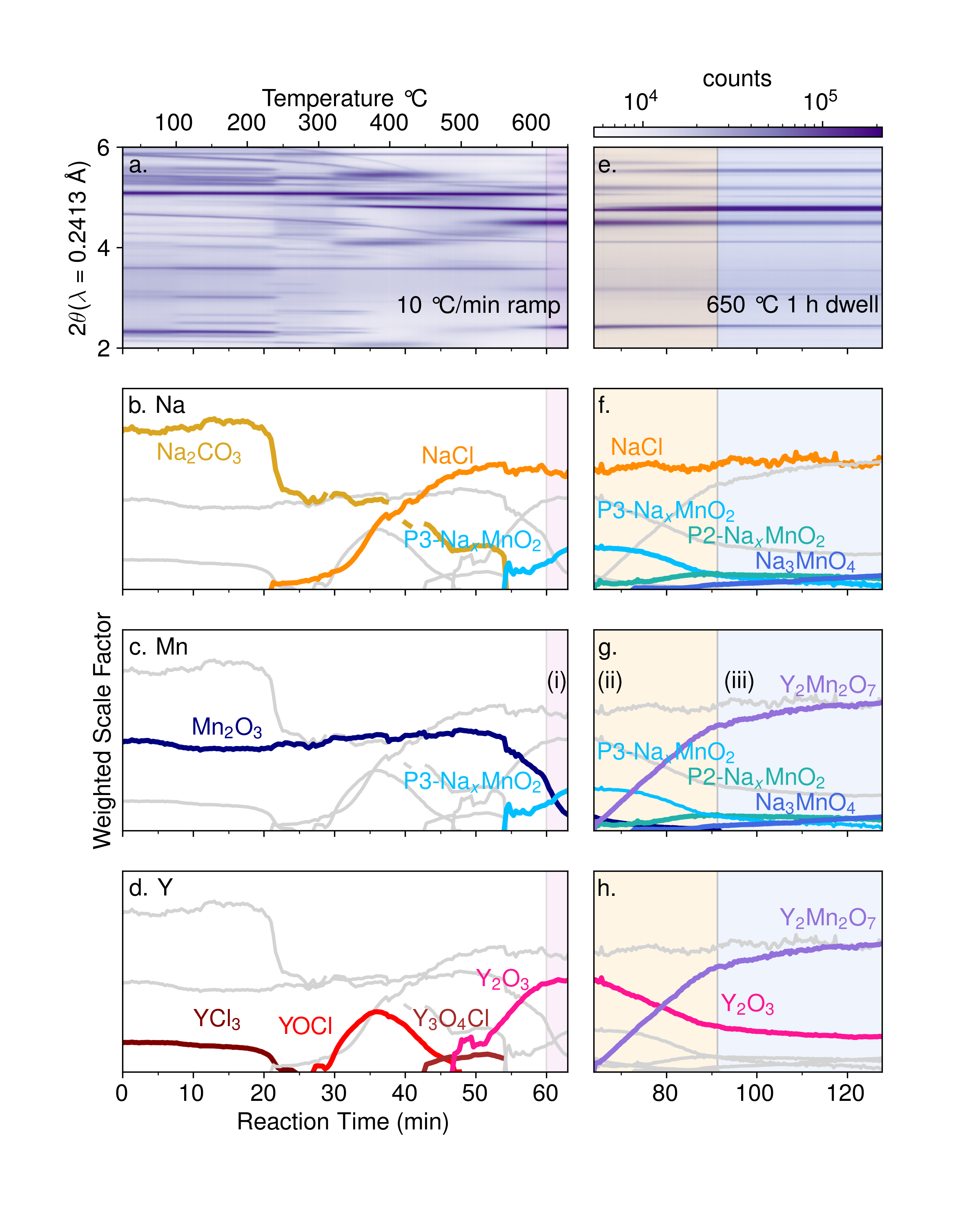}
\caption{\label{fig:insitucationplot} 
\textbf{\textit{In situ} SPXRD data and calculated weighted scale factors from sequential Rietveld analysis as a function of time for the reaction: \ce{Mn2O3 + 2\text{~}YCl3 + 3\text{~}Na2CO3 + 1/2\text{~}O2 -> Y2Mn2O7 + 6\text{~}NaCl + 3\text{~}CO2} under flowing oxygen}. The reaction is separated into two main vertical panels: (a-d) heating at 10 \textcelsius{}/min to 650\textcelsius{} and (e-h) dwelling at 650 \textcelsius{} for 60 min. \textbf{a,e}, The \textit{in situ} diffraction data plotted using a false color representation of the diffraction counts. Calculated weighted scale factors of phases over the course of the reaction are plotted and separated into horizontal panels by cation element: \textbf{b,f}, sodium, \textbf{c,g}, manganese, and \textbf{d,h}, yttrium. The gray lines in each horizontal panel show the observed phases containing the other cations. The shaded regions, (i), (ii), and (iii), correspond to trajectories highlighted in Figure~\ref{fig:fstarP3}.}
\end{center} 
\end{figure}

\textit{In situ} synchrotron powder X-ray diffraction (SPXRD) studies were used to identify the intermediate phases and reaction pathways that permit the selective formation of \ce{Y2Mn2O7} through assisted metathesis reactions with \ce{Na2CO3}. Figure~\ref{fig:insitucationplot} shows the integrated SPXRD diffraction patterns collected during a reaction using precursor phases of \ce{Mn2O3 + 2YCl3 + 3Na2CO3}. Data were collected during heating to 650 \textcelsius{} at 10 \textcelsius{}/min then dwelling at 650 \textcelsius{} for 60 min, performed under flowing oxygen. Quantitative phase analysis using the Rietveld method reveals the identities and phase fractions of intermediates of P3-\ce{Na_$x$MnO2}, P2$^\prime$-\ce{Na_$x$MnO2}, YOCl, \ce{Y3O4Cl}, \ce{Y2O3}; products of \ce{NaCl} and \ce{Y2Mn2O7}; and a small amount of \ce{Na3MnO4} impurity, as summarized in Figure~\ref{fig:insitucationplot}. No other yttrium manganese oxide phases were observed.

The relevant intermediates that exist at the onset of formation of \ce{Y2Mn2O7} are \ce{Y2O3}, P3-\ce{Na_{$x$}MnO2}, and a small amount of \ce{Mn2O3}. \ce{Y2O3} forms with a very small particle size (less than $\sim$35 nm, Figure~\ref{S-fig:particlesizeY2O3}), and its production yields a large fraction of the expected NaCl product (Figure~\ref{fig:insitucationplot}). P3-\ce{Na_{$x$}MnO2} forms directly from the reaction of \ce{Na2CO3} and \ce{Mn2O3}, in agreement with a previous \textit{in situ} and computational study.\cite{Bianchini2020} Once both of these key intermediates are present, P3-\ce{Na_{$x$}MnO2} then reacts with \ce{Y2O3} to form pyrochlore \ce{Y2Mn2O7}, also triggering the formation of P2$^\prime$-\ce{Na_{$x$}MnO2} ($Cmcm$ structure, with a slight shear of the ideal P2-$P6_{3}/mmc$ structure)\cite{Caballero2002, Co1999, Clement2015}. Thereafter, selective formation of \ce{Y2Mn2O7} is sustained through the reaction of P2$^\prime$-\ce{Na_{x}MnO2} with \ce{Y2O3}. After P3-\ce{Na_{$x$}MnO2} is fully consumed, the rate of \ce{Y2Mn2O7} production slows dramatically, as \ce{Na3MnO4} gradually forms (Fig~\ref{fig:insitucationplot}). \ce{Na3MnO4} does not appear in reaction products of \textit{ex situ} studies of these assisted metathesis reactions performed on bulk scales, although it may be present as a trace quantity.\cite{Todd2019} 

\subsection{Thermodynamic stability of \ce{Y2Mn2O7}}

\begin{figure}[ht!]
\begin{center}
\includegraphics[width=6in]{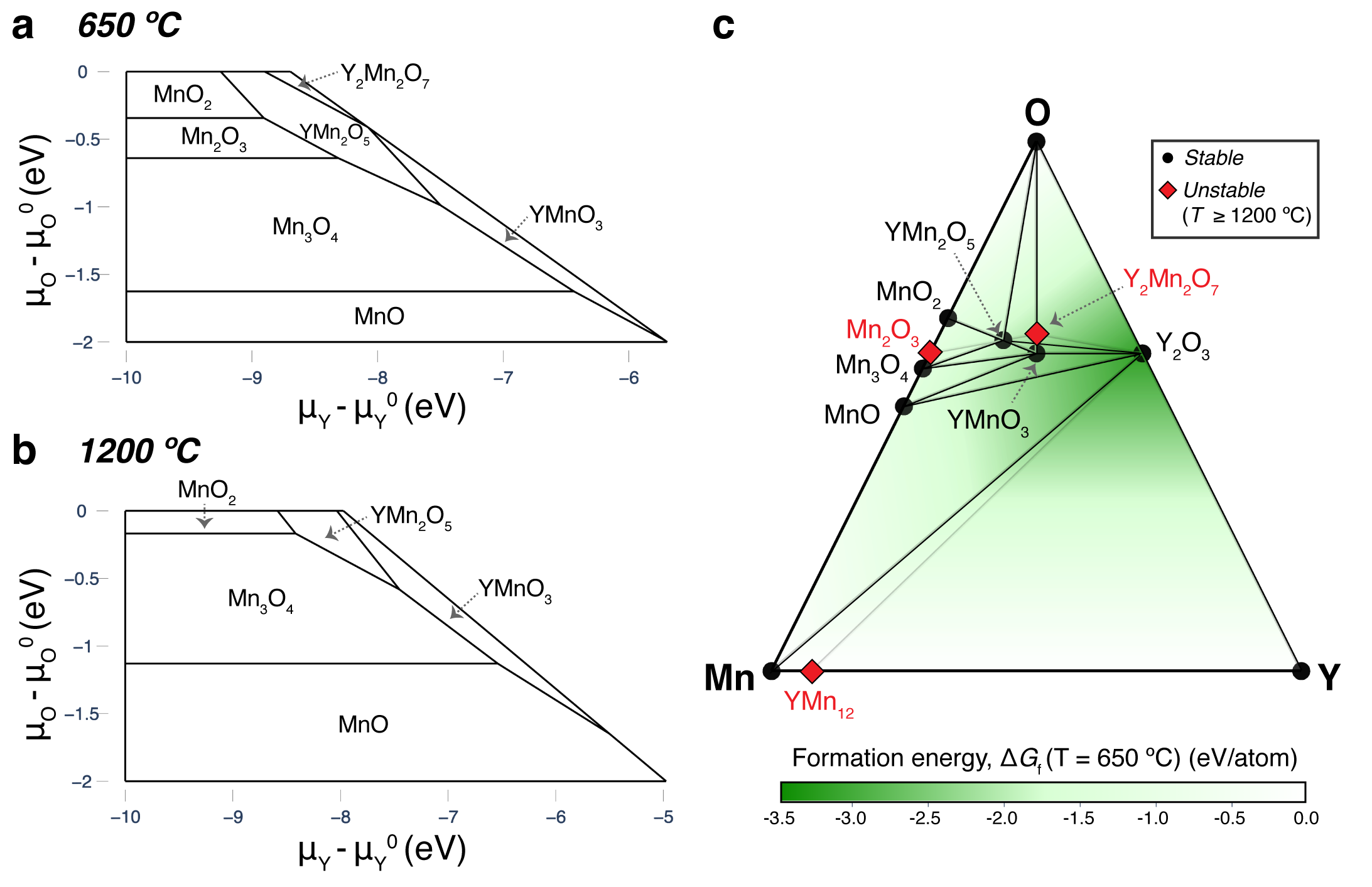}
\caption{\label{fig:YMnOPhaseDiagram} \textbf{Computed phase stabilities in the Y-Mn-O chemical system.} Predominance diagrams are shown as a function of oxygen and yttrium chemical potentials, $\mu_{\text{O}}$ and $\mu_{\text{Y}}$, at temperatures of \textbf{a}, 650 \textcelsius{} and \textbf{b}, 1200 \textcelsius{}, referenced to their standard elemental state. \textbf{c}, Predicted ternary compositional phase diagram for the Y-Mn-O system at 1200 \textcelsius{} (black lines/circles) overlaid on the phase diagram at 650 \textcelsius{}, where red squares and gray lines mark phases and facets that are destabilized at T $\geq$ 1200 \textcelsius{}, respectively. \ce{Y2Mn2O7} is predicted to be stable at low temperatures ($T \leq$ 1100 \textcelsius{}). Gibbs free energies of formation are estimated by applying a machine-learned transformation \cite{Bartel2018} on DFT-based formation enthalpies acquired from the Materials Project database.\cite{Jain2013}.}
\end{center} 
\end{figure}

Evaluation of the Y-Mn-O phase diagram reveals that \ce{Y2Mn2O7} is thermodynamically stable (i.e., it is on the convex hull) at low temperatures (T $\leq$ 1100 \textcelsius{}). Using error-corrected formation energies derived from DFT calculations \cite{Jain2011} and a previous Gibbs free energy model for solids \cite{Bartel2018}, the phase diagram of the Y-Mn-O system has been calculated at finite temperatures relevant to the synthesis reactions being investigated, with results of calculations at 650 \textcelsius{} and 1200 \textcelsius{} shown in Figure~\ref{fig:YMnOPhaseDiagram}. Figure~\ref{fig:YMnOPhaseDiagram}a depicts that the predominance area of \ce{Y2Mn2O7} is quite small, indicating its low relative stability with respect to thermodynamic decomposition to neighboring phases. For a 1:1 Y:Mn composition ratio in an environment open to oxygen, h-\ce{YMnO3} is thermodynamically stable at all temperatures and under a wide range of oxygen chemical potentials, while \ce{Y2Mn2O7} is thermodynamically stable only below $\sim$1100 \textcelsius{} and at higher oxygen chemical potentials (i.e., oxygen rich conditions). While the predicted temperature below which \ce{Y2Mn2O7} is stable is overestimated by these calculations since only h-\ce{YMnO3} was observed to form at 950 \textcelsius{} in experiments \cite{golikov}, the low-temperature thermodynamic stability of \ce{Y2Mn2O7} is assessed experimentally here by heating \ce{Y2Mn2O7} in flowing oxygen at 650 \textcelsius{} for two weeks (Figure~\ref{S-fig:Y2Mn2O7}), a result that is in agreement with previous thermogravimetric analysis \cite{Subramanian1988}. The fact that the \ce{Li2CO3} and \ce{K2CO3} precursors do not produce the thermodynamically favored phase of \ce{Y2Mn2O7} at temperatures below 950 \textcelsius{} and $p$(\ce{O2}) = 1 atm  (i.e., $\mu_O-\mu_O^0$ $\approx$ 0) indicates that the metathesis reaction temperatures are sufficiently low to yield kinetic control in the formation of o-\ce{YMnO3} or h-\ce{YMnO3} \cite{Martinolich2016b}. Finally, it indicates the need for a deeper understanding of the specific factors that determine which reaction pathway is followed for a given alkali carbonate precursor.

\subsection{Product selectivity via chemical potential distances}

\begin{figure}[ht!]
\begin{center}
\includegraphics[width=6.5in]{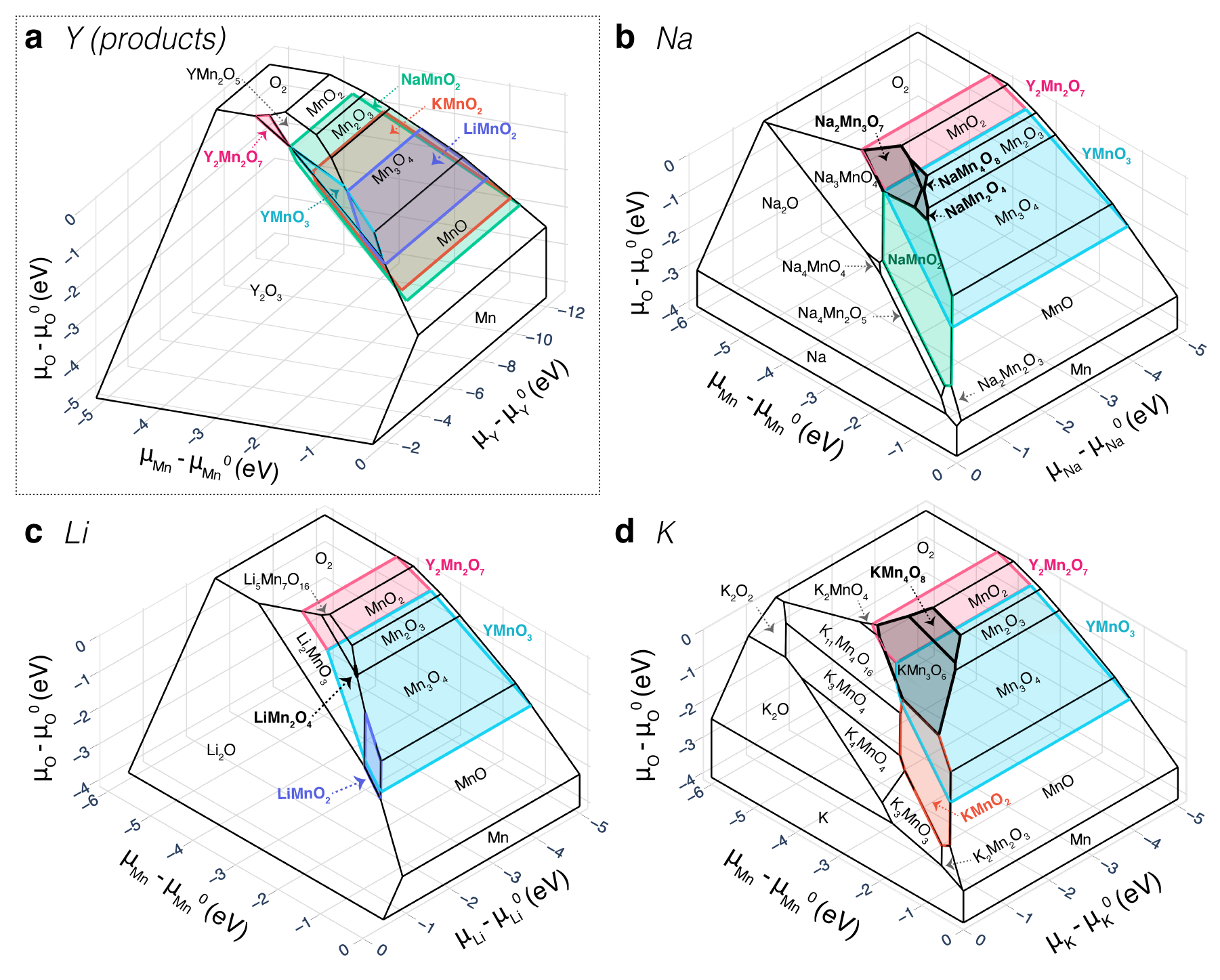}
\caption{\label{fig:wormholes} \textbf{Chemical potential diagrams of the Y-Mn-O-Na-Li-K system at 650 \textcelsius{}, plotted in product (boxed) and reactant subspaces}. \textbf{a}, Y-Mn-O product subspace where the stability polytopes of stoichiometric \ce{NaMnO2} (green), \ce{LiMnO2} (purple), and \ce{KMnO2} (orange) have been plotted as lower-dimensional slices, appearing as thin polyhedra due to the added degree of freedom ($\mu_\text{Y}$). The \ce{NaMnO2} polyhedron extends to high enough $\mu_\text{O}$ and low enough $\mu_\text{Mn}$ to reach \ce{Y2Mn2O7}, while the \ce{LiMnO2} and \ce{KMnO2} polyhedra do not. As such, \ce{LiMnO2} and \ce{KMnO2} suggest the formation of \ce{YMnO3} and/or \ce{YMn2O5} depending on the value of $\mu_\text{Y}$, as previously observed.\cite{Todd2019,Todd2019a,Todd2020}. \textbf{b-d}, A-Mn-O reactant subspaces with visualized slices of the \ce{YMnO3} (cyan) and \ce{Y2Mn2O7} (magenta) polytopes. The other shaded areas (gray) highlight \ce{A_{$x$}MnO2} intermediates (and structurally-related neighbors) that lead directly to the formation of yttrium manganese oxide products.}
\end{center} 
\end{figure}

We hypothesize that the generalized chemical potential diagram, which was previously devised as a three-dimensional extension of predominance diagrams with application to modeling interface stability and reactivity \cite{Yokokawa1999}, clarifies the large differences in product selectivity that arise by changing the identity of the alkali precursor. As the mathematical dual to the convex hull of the extensive energy-composition space, the generalized chemical potential diagram explicitly defines phase stability regions as bounded convex $n-1$ dimensional polytopes within the full $n$-dimensional, intensive chemical potential space. Although in principle the chemical potential diagram conveys identical information as the traditional compositional phase diagram, it has the unique advantage of directly revealing the \textit{relative} stability of phases. In fact, the volume of each phase's stability polytope increases with its energy ``below'' the convex hull, i.e., the energy that would be released forming this phase via decomposition from the neighboring phases within a facet of the hull. The chemical potential diagram also directly links thermodynamic phase construction with atomic diffusion, and its construction even permits the visualization of diffusion paths in solid-state reactions \cite{Yokokawa1999}. Hence the chemical potential diagram is well-suited for application to understanding and predicting the local behavior of reactions at solid interfaces.

Selective reactions in solid state chemistry need to minimize the probability of forming an undesired product phase at any point in time during the reaction pathway. As has been previously proposed,\cite{Schmalzried1981} a reaction at the interface between two solid phases can proceed through either interface control or diffusion control, the latter resulting in local thermodynamic equilibrium and the corresponding requirement that elemental chemical potentials be continuous across the interfaces between reactants and product phase(s). All solid-state reactions eventually tend towards diffusion control as the product layer thickness grows. For this reason, predicted reactions which pose synthetic challenges are those in which it is impossible to achieve local equilibrium without decomposition to other phases before formation of the equilibrium product. On the chemical potential diagram, such reactions involve reactant-product pairs which do not share a phase boundary. Considering the reaction \ce{Y2O3 + Mn2O3 -> 2\text{~}YMnO3}, if local equilibrium is achieved during the reaction, then according to Figure~\ref{fig:wormholes}a, one expects \ce{YMn2O5} and/or \ce{Mn3O4} to form before formation of the final \ce{YMnO3} product, consistent with the results of control reactions presented in Figure~\ref{S-fig:MnOControls} and the previously-reported phase diagram \cite{golikov}. 

\begin{figure}[ht!]
\begin{center}
\includegraphics[width=6.5in]{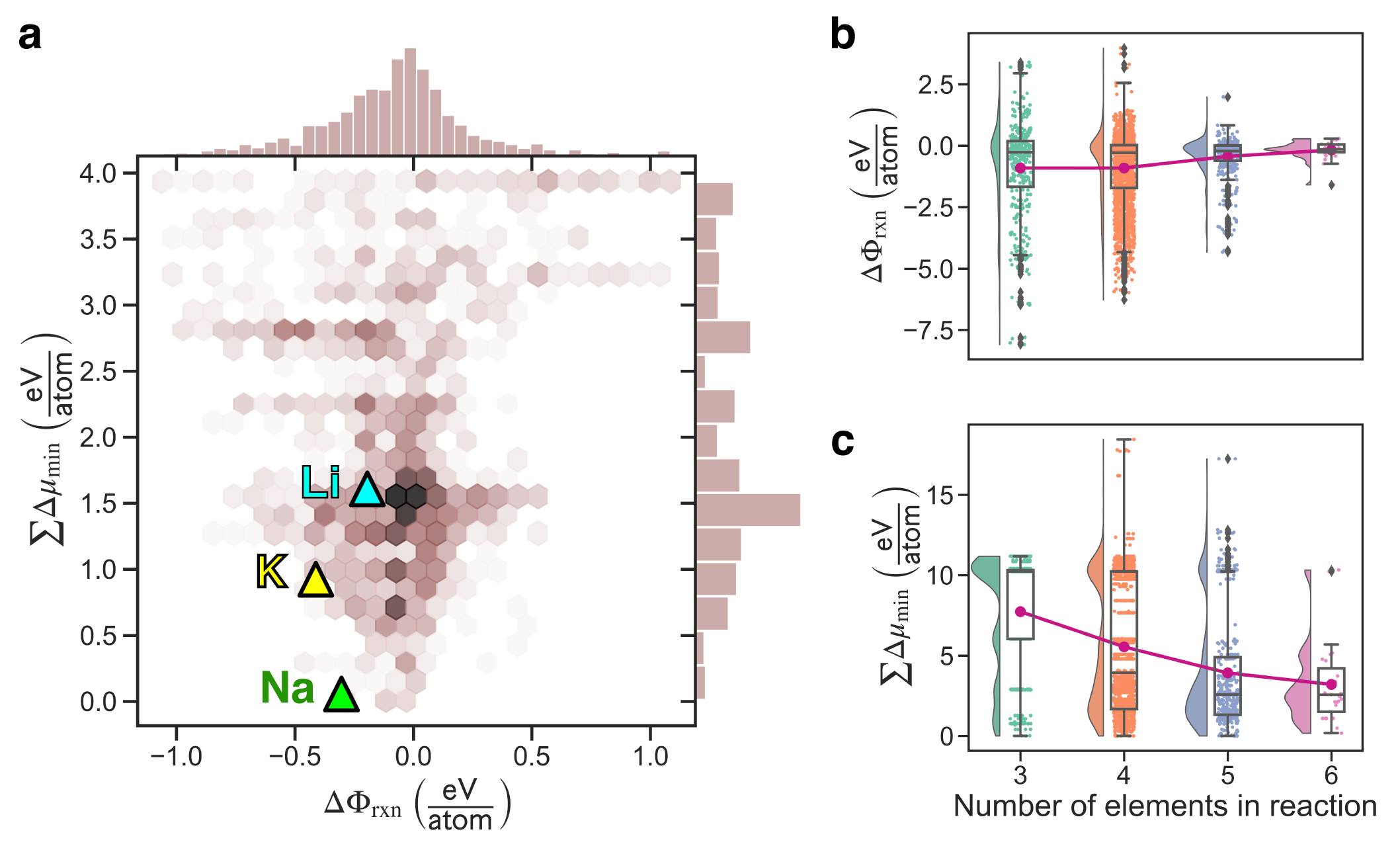}
\caption{\label{fig:y2mn2o7_rxns} \textbf{Energy and total chemical potential distance distributions for 3,017 predicted chemical reactions forming \ce{Y2Mn2O7}}. The reactions were predicted from a chemical space consisting of all alkali metals (Li, Na, K, Rb, Cs), halogens (F, Cl, Br, I), carbon (C), and the target elements (Y, Mn, O). Oxygen was treated as an open element with a chemical potential corresponding to $\mu_\mathrm{O}^0$ ($p=0.1$ MPa, $T=650$ \textcelsius{}). \textbf{a}, A hexbin plot showing the distribution of reaction energies (in grand potential), $\Delta \Phi_{\mathrm{rxn}}$, and calculated total chemical potential distance values, $\sum\Delta \mu_{\mathrm{min}})$ for each reaction. The three colored triangles indicate the specific reactions \ce{2\text{~}$A$MnO2 + 2\text{~}YOCl + 1/2\text{~}O2 -> Y2Mn2O7 + 2\text{~}$A$Cl} for $A$=Li, Na, K. \textbf{b,c}, Distributions of reaction energies and total chemical potential distances as a function of the number of elements in the reaction. All reactions include at least the three elements of the target phase (Y, Mn, O). The mean values for each distribution are connected by a magenta line.}
\end{center} 
\end{figure}

Therefore, selective reactions are those which can be balanced, produce the desired product, and minimize the distance in chemical potential space between pairwise reactant-product interfaces.  On the chemical potential diagram, this manifests as determining whether each reactant-product pair shares a boundary (i.e., zero distance) or is separated by the polytope(s) of one or more additional phases (i.e., non-zero distance). This also applies to product-product pairs, which may further react if they do not share a boundary on the chemical potential diagram. Plotting trial reactants in the three-dimensional chemical potential subspace of the product (Figure~\ref{fig:wormholes}a), or plotting products in the reactant subspace (Figure \ref{fig:wormholes}b-d), permits direct visualization of the pairwise interfaces that control reactivity \cite{Miura2020b,Bianchini2020}.  This ``chemical potential distance", $\Delta \mu_{\mathrm{min}}$ is geometrically calculated for all reactant-product and product-product interfaces in a reaction by determining the minimum Euclidean distance between polytopes (see Methods). The primary advantage of calculating a distance value rather than a Boolean variable for whether or not two phases share a boundary is that the distance is more robust against inaccuracies in thermochemical data and conveys the degree to which the competing phases may interfere with the selectivity of the reaction. 

Visualizing three-dimensional slices of the full Y-Mn-O-Na-Li-K chemical potential space reveals why only the sodium-based intermediates lead to formation of \ce{Y2Mn2O7}. Figure~\ref{fig:wormholes}a illustrates slices of the stoichiometric \ce{AMnO2} intermediates in the Y-Mn-O product chemical potential space. The \ce{NaMnO2} stability polyhedron extends well beyond the range of both the \ce{LiMnO2} and \ce{KMnO2} phases. In fact, this Na-containing phase extends into high enough $\mu_\text{O}$ values to reach the stability area for \ce{Y2Mn2O7}. This enhanced stability of the \ce{NaMnO2} phase into higher oxygen (and lower manganese) chemical potentials suggests that it may be uniquely capable of forming a stable interface with \ce{Y2Mn2O7} and facilitating direct reaction to this pyrochlore phase. Specifically, the nearly shared boundary between the \ce{NaMnO2} intermediate and \ce{Y2Mn2O7} means that the reaction kinetics should be facile since the local chemical potentials of Mn and O do not need to change as the intermediate converts to the product. Furthermore, even though the \textit{global} oxygen chemical potential is controlled through partial pressure of \ce{O2}, the reaction of \ce{NaMnO2} provides \textit{locally} available oxygen at the chemical potential required to form \ce{Y2Mn2O7}. In the case of the A = Li, K precursors, the lack of available oxygen with an appropriately high chemical potential necessitates the reaction to proceed through other intermediates before it is even possible to reach \ce{Y2Mn2O7}.

The complementary inverse slices of the Y-Mn-O ternaries within the reactant A-Mn-O spaces (Figure~\ref{fig:wormholes}b-d), along with the study of the atomic structures of the \ce{AMnO2} phases, provide further explanation as to why the sodium-based intermediates facilitate reaction to \ce{Y2Mn2O7} even for Na-deficient phases, \ce{Na_{$x$}MnO2} (where x $<$ 1). In agreement with prior work \cite{Kitchaev2017}, the sodium-based phases show a strong preference for the layered $\alpha$-NaFeO$_2$-type structures relative to tunnel-like (e.g., Ramsdellite-derived) or spinel-derived structures. \ce{Na2Mn3O7}, \ce{NaMn2O4}, and \ce{NaMn4O8}, illustrated as gray shaded areas in Figure~\ref{fig:wormholes}b, are all predicted to share the structurally analogous layered framework with stoichiometric \ce{NaMnO2} (Figure~\ref{S-fig:structures}a). Together these structures create an interconnected pathway along the Na-Mn-O chemical potential surface towards higher oxygen chemical potentials where the surface touches the \ce{Y2Mn2O7} product (e.g., oxidation states ranging from +3 to +5 depending on the nature of the vacancies and whether the composition is Na-poor or Mn-poor).  In contrast, \ce{LiMnO2} has a narrower range of stability than its neighboring Li-Mn-O phases, which each prefer distinct non-layered structures (Figure~\ref{S-fig:structures}b). In both the \ce{Li2CO3} assisted metathesis reactions \cite{Todd2019a} and the direct reaction between \ce{LiMnO2} and \ce{YOCl} \cite{Todd2020}, we observed formation of \ce{YMnO3} and \ce{YMn2O5}, suggesting that the decreased range of chemical potentials over which layered \ce{LiMnO2} is stable and lack of structural homology between the \ce{Li_{$x$}MnO2} phases inhibits the accessibility of \ce{Y2Mn2O7} (e.g., due to rearrangement of the phases into spinel-derived structures). Finally, in the \ce{K2CO3} assisted metathesis reactions, no product selectivity was observed as the reaction produced a mixture of \ce{YMnO3}, \ce{YMn2O5}, and \ce{Y2Mn2O7} products.\cite{Todd2019a} While the \ce{K_{$x$}MnO2} phases indeed exhibit a similar extension into high oxygen chemical potentials akin to \ce{Na_{$x$}MnO2} (Figure~\ref{fig:wormholes}d), the stoichiometric \ce{KMnO2} ground-state structure does not follow the layered (nor even another another common) \ce{MnO2} framework,\cite{Kitchaev2017} (Figure~\ref{S-fig:structures}c) although the layered phase is predicted to be about 44 meV/atom higher in energy than the ground state. We thus conclude that similar to Li, the K-based system lacks the structural connectivity for interconversion between different \ce{K_{$x$}MnO2} intermediates, although direct formation of K-deficient phases may promote formation of \ce{Y2Mn2O7}. In both the Li and K systems, the initial formation of \ce{YMnO3} or \ce{YMn2O5} hence imposes significant kinetic barriers to the formation of \ce{Y2Mn2O7}. According to the chemical potential diagrams, \ce{YMnO3} would have to decompose into \ce{YMn2O5} and \ce{Y2O3} en route to forming \ce{Y2Mn2O7}, which is consistent with myriad control experiments between binary yttrium and manganese oxides, with or without the presence of sodium (see Figures~\ref{S-fig:MnOControls} and \ref{S-fig:MnONaClControls}). Based on these observations, we argue that the minimized chemical potential distance presented in Figures~\ref{fig:wormholes} and \ref{fig:y2mn2o7_rxns} and ability of sodium-based intermediate to interconvert across a wide range of stoichiometries and oxidation states explain why sodium provides selectivity of \ce{Y2Mn2O7} via a kinetically-viable pathway at lower temperatures.

Analysis of 3,017 unique chemical reactions predicted to form \ce{Y2Mn2O7} within the full alkali metal (Li, Na, K, Rb, Cs), halogen (F, Cl, Br, I), carbon (C), and target (Y, Mn, O) chemical system show that the Na-based ternary metathesis reaction, \ce{2\text{~}NaMnO2 + 2\text{~}YOCl + 1/2\text{~}O2 -> Y2Mn2O7 + 2\text{~}NaCl} is the most optimal reaction for both minimizing the total chemical potential distance and maximizing the (negative) grand potential energy of reaction, $\Delta \Phi_{\mathrm{rxn}}$, when these two parameters are weighted equally (see Methods). This result suggests that the Na-based pathway is not only a good pathway for producing \ce{Y2Mn2O7} but that it may also be the most suitable metathesis reaction for making this product. The full set of reactions and their weighted rankings can be found in provided data in the Supporting Information. The energy and total chemical potential distributions for these reactions illustrate that the Na-based reaction (green triangle in Figure \ref{fig:y2mn2o7_rxns}a) optimizes both total chemical potential distance and overall reaction energy. The Na-based reaction has a near-zero total chemical potential distance (0.055 eV/atom) that is significantly  smaller than those of the Li and K reactions (1.608 and 0.924 eV/atom respectively)  and has a more negative reaction energy than any other reaction pathway with a near-zero total chemical potential distance. Figure \ref{fig:y2mn2o7_rxns} also reveals a major design principle behind using an expanded chemical space with metathesis reactions -- the addition of other elements beyond those of the target phase (Y, Mn, O) can decrease the mean total chemical potential distance of the reaction without significantly changing the mean reaction energy. This hence provides many more opportunities for finding selective reactions beyond those that exclusively contain the target chemical species, i.e., 3,017 reactions in the full 13-element chemical system vs. 362 potential reactions in the Y-Mn-O chemical system.

Direct ternary metathesis reactions carried out between \ce{Na_$x$MnO2} and \ce{YOCl} do indeed form \ce{Y2Mn2O7} as predicted, but they also reveal important mechanistic details missing from the presented thermodynamic analysis that impacts product selectivity.  For example, available density functional theory data is highly focused on stoichiometric compounds but tends to be lacking for solid solution and defect-containing phases. The reaction \ce{NaMnO2 + YOCl ->[O$_{2}$]} results in some \ce{Y2Mn2O7} formation, but also the formation of \ce{YMnO3} and \ce{YMn2O5}  (Figure~\ref{S-fig:mNaMnO2YOCl}), suggesting that the evolution of the \ce{Na_$x$MnO2} precursor during the reaction is important to consider, as illustrated in Figure~\ref{fig:wormholes}b. A partially-oxidized precursor in the reaction, P2$^\prime$-\ce{Na_{0.7}MnO2} + YOCl selectively yields \ce{Y2Mn2O7}, but the reaction is incomplete after 24 h (Figure~\ref{S-fig:oNaMnO2YOCl}). Together, these control reactions suggest that the reaction rates also depend on the specific nature of defects present in the reactants and intermediates.

\subsection{Mechanistic facilitation through defect reactions}

\begin{figure}[ht!]
\begin{center}
\includegraphics[width=5in]{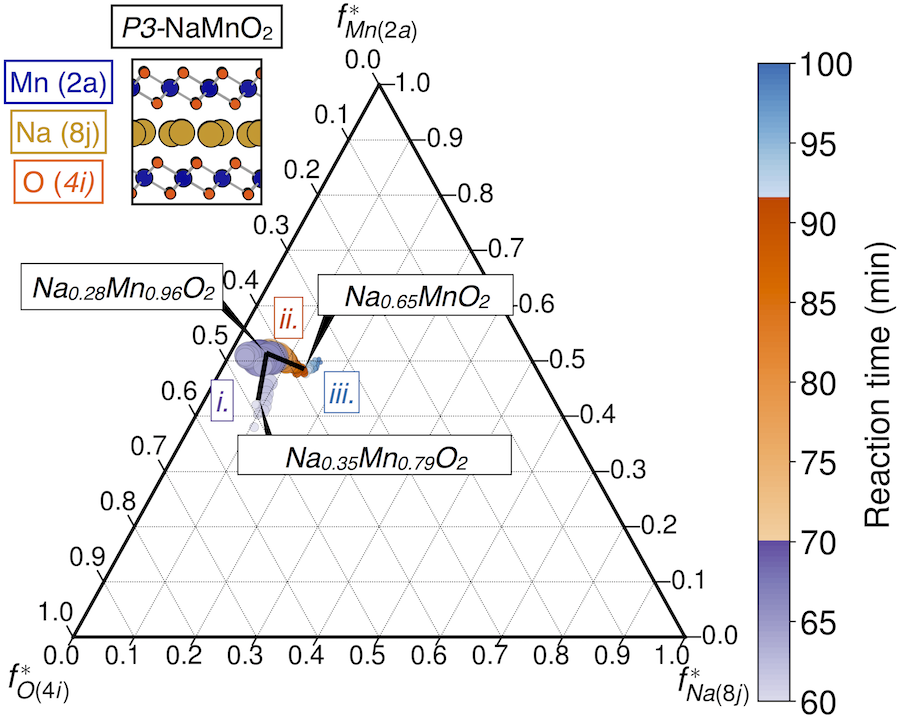}
\caption{\label{fig:fstarP3} \textbf{Compositional variation of the P3-\ce{Na_{$x$}MnO2} intermediate determined by Rietveld analysis.}  In the $f^{*}$ diagram, each axis denotes the X-ray scattering power from a crystallographic Wyckoff site within the lattice. For reference, calculated trajectories that correspond to the different stoichiometries of P3-\ce{Na_{$x$}MnO2} are provided as black lines. The color bar in the figure follows the evolution of the P3-\ce{Na_{$x$}MnO2} intermediate in the assisted metathesis reaction presented in Figure~\ref{fig:insitucationplot} with highlighted trajectories (i-iii) reflecting shaded regions in Figure~\ref{fig:insitucationplot}(i-iii). The diameter of each circle mirrors the calculated weighted scale factor (WSF) in Figure~\ref{fig:insitucationplot} for P3-\ce{Na_{$x$}MnO2} and thus corresponds to the amount of this layered phase. The structure of P3-\ce{Na_{$x$}MnO2} is shown with Wyckoff atom sites labeled.  The figure contains compositions for which the WSF is greater than 20\% of the maximum WSF for the phase shown in Figure~\ref{fig:insitucationplot}. Detailed analysis of individual site occupancies is provided in the SI.}
\end{center} 
\end{figure}

As the reaction pathway of the Li-based assisted metathesis reaction was previously reported \cite{Todd2019a}, we focus on how \ce{Na_{$x$}MnO2}-based intermediates react in a manner consistent with the predicted thermodynamic connectivity. Crystallographic analysis of the \ce{Na_{$x$}MnO2} phases by free refinement of the crystallographic site occupancies of all elements in each \ce{Na_{$x$}MnO2} structure and their subsequent analysis on an $f^*$ diagram\cite{Yin2018} together reveal how the stoichiometry of  \ce{Na_{$x$}MnO2}  changes through three different defect reactions along the reaction pathway for P3-\ce{Na_{$x$}MnO2} (Figure~\ref{fig:fstarP3}) and for P2$^\prime$-\ce{Na_{$x$}MnO2} (Figure~\ref{S-fig:fstarP2}). While the exact compositions defined by the site occupancies are correlated with other refinement variables (e.g., atomic displacement parameters, as addressed in the Supporting Information), the compositional trajectories observed in Figure~\ref{fig:fstarP3} are robust\cite{Yin2018}. Two distinct compositional trajectories of the P3-\ce{Na_{$x$}MnO2} phase on Figure~\ref{fig:fstarP3} follow two major processes during the reaction pathway: (i) P3-\ce{Na_{$x$}MnO2} formation and (ii) reaction of P3-\ce{Na_{$x$}MnO2} to yield \ce{Y2Mn2O7} and P2$^\prime$-\ce{Na_{$x$}MnO2}. Additionally, delineation of trajectory (iii) describes when P2$^\prime$-\ce{Na_{$x$}MnO2} begins reacting to yield \ce{Y2Mn2O7}. 

The reaction of the P3-\ce{Na_{$x$}MnO2} follows a compositional trend in chemical potential space towards the \ce{Y2Mn2O7} product (Figure~\ref{fig:wormholes}).  The initial formation of P3-\ce{Na_{$x$}MnO2} results in a very sodium-deficient composition, P3-\ce{Na_{0.29}Mn_{0.96}O2}. This manifests crystallographically as an excess of scattering intensity on the sodium site relative to the manganese site, suggestive of anti-site disorder,  as well as decreased scattering relative to oxygen, suggestive of metal deficiency (Figure~\ref{fig:fstarP3}, as described in detail in the Supporting Information), which is also observed in the initial formation of \ce{LiNiO2} \cite{Bai2020}. Additionally, there is decreased scattering intensity of the metal sites relative to oxygen, which suggests the existence of overall cation deficiencies in the structure, calculated as \ce{Na_{0.35}Mn_{0.79}O_{2}} from Rietveld analysis of the PXRD data. The composition evolves in time through a cation-ordering reaction, where manganese occupying the ($8j$) sodium site moves to the manganese ($2a$) site (trajectory (i) in Figure~\ref{fig:fstarP3}). Movement along trajectory (i) yields site ordering as the phase fraction is increased; the proposed stoichiometry is analytically calculated as the black line overlaid on these data in Figure~\ref{fig:fstarP3} (see Eqn.~\ref{S-eq:cationswap}). This ordering reaction yields a calculated composition of P3-\ce{Na_{0.29}Mn_{0.96}O2} at its maximal phase fraction observed in Figure~\ref{fig:insitucationplot}. At that point in the overall reaction, this intermediate with a relatively high oxygen chemical potential (Figure~\ref{fig:wormholes}b) reacts to form \ce{Y2Mn2O7}.

There is a change in the defect chemistry of P3-\ce{Na_{$x$}MnO2} consistent with the loss of Mn$^{4+}$ + 2 O$^{2-}$, concomitant with consumption of \ce{Y2O3} and production of \ce{Y2Mn2O7} and P2$^\prime$-\ce{Na_{$x$}MnO2}. Along trajectory (ii) in  Figure~\ref{fig:fstarP3}, there is a decrease in relative electron density from the oxygen site ($4i$), as well as an increase in the sodium to manganese site ratio, as described by Eqn.~\ref{S-eq:simulreactions}. This reflects a change in defect chemistry along trajectory (ii) from the P3-\ce{Na_{0.29}Mn_{0.96}O2} calculated composition at the end of trajectory (i) to that of P3-\ce{Na_{0.65}MnO2}.  During this process, the pyrochlore phase fraction grows rapidly and P2$^\prime$-\ce{Na_{$x$}MnO2} forms (Figure~\ref{fig:insitucationplot}, at $t=$ 70--90 min) before tapering off at the end of the second trajectory in Figure~\ref{fig:fstarP3} as the amount of P3-\ce{Na_{$x$}MnO2} depletes. 

Analysis of the \textit{in situ} diffraction data reveals that selective formation of \ce{Y2Mn2O7} occurs through two parallel reactions that start during trajectory (ii) in Figure~\ref{fig:fstarP3}:
\begin{subequations}
\begin{align}
\label{eq:elementary1}
\cee{4\text{~P3-}Na_{0.25}MnO2 + Y2O3 & -> 2 \text{~P2$^\prime$-}Na_{0.5}MnO2  +  Y2Mn2O7} \\
\label{eq:elementary2}
\cee{12\text{~P2$^\prime$-}Na_{0.5}MnO2 + 8\text{~}O2 + 5\text{~}Y2O3  & -> 5\text{~}Y2Mn2O7 + 3\text{~}Na3MnO4}, 
\end{align}
\end{subequations}
consistent with the predicted thermodynamic connectivity in Figure~\ref{fig:wormholes}.  In contrast, Li-based assisted metathesis reactions performed at temperatures between 500 \textcelsius{} and 850 \textcelsius{} proceed via \ce{LiMnO2}-based intermediates that result in the direct formation of \ce{YMnO3} phase \cite{Todd2019a,Todd2020}, which is also predicted from the thermodynamic connectivity (Figure~\ref{fig:wormholes}a,c).  

Based on the stoichiometry of the residual Na-Mn-O phases in each step, there is a loss of Mn$^{4+}$ + 2 O$^{2-}$ along with the consumption of \ce{Y2O3}. Thermodynamically, we expect that stoichiometric \ce{NaMnO2} reacts with \ce{Y2O3} at a relatively high $\mu_\text{O}$ and low $\mu_\text{Mn}$ (see Figure~\ref{fig:wormholes}b), as suggested by the average stoichiometry of the transferred species in the parallel cascade of defect reactions (Eqns.~\ref{eq:elementary1} and \ref{eq:elementary2}). Curiously, the \ce{Y2Mn2O7}-forming reaction does not proceed via a typically labile topochemical deintercalation of the alkali cation \cite{Si2012} or oxygen anion  \cite{Hayward1999,Tsujimoto2007,Yajima:2015db}, but instead the chemical potential boundaries shown in Figure~\ref{fig:wormholes} guide the reaction to direct formation of the \ce{Y2Mn2O7} product from the Na-Mn-O intermediates by avoiding the formation of other Y-Mn-O phases.

\section{Conclusions}

In assisted metathesis reactions, the presence of an appropriate alkali ion controls the selectivity of the reaction via differences in the thermodynamic stabilities of the reaction intermediates, as defined by the chemical potentials of the constituents in a hyperdimensional compositional space. In the reaction of \ce{3Na2CO3 + 2YCl3 + Mn2O3}, the \ce{Y2Mn2O7} pyrochlore forms selectively via the direct reaction of \ce{Na_{$x$}MnO2} and \ce{Y2O3} intermediates, as revealed by \emph{in situ} temperature- and time-dependent SPXRD experiments. However, this pyrochlore selectivity is not achieved with the analogous \ce{Li2CO3} or \ce{K2CO3} precursors. Mapping the reaction species in chemical potential space illustrates that \ce{NaMnO2}, and structurally-related compositional variants permit a small difference in local chemical potentials between the Na$_x$Mn$_y$O$_2$ intermediates and \ce{Y2Mn2O7}, as the chemical stability windows of structurally homologous Na$_x$Mn$_y$O$_2$ phases connect directly with that of \ce{Y2Mn2O7}. The short distance in chemical potential space illustrates why the reaction to form \ce{Y2Mn2O7} is selective. In contrast, the other alkali-based (Li, K) intermediate phases lack the same degree of structural homology and share different boundaries, thus resulting in the formation of different products. The mechanistic details obtained from crystallographic analysis highlight that the thermodynamic boundaries dictate the reactivity rather than the mobility of an alkali cation within the oxide framework. This analysis illustrates how local thermodynamic equilibrium principles provide a direct connection to reaction kinetics in guiding the mechanistic pathway of solid-state reactions, thus providing a notion of a protecting group in materials synthesis. The thermodynamic underpinnings permit high-throughput prediction of precursors in higher dimensional chemical spaces that can impart selectivity, kinetic viability, and reversibility.

\section{Supporting Information}
\begin{itemize}
    \item Figures S1-\ref{S-fig:pareto}: Assisted metathesis reaction energetics, detailed description of the Na-based reaction progression, testing of alternative \ce{NaMnO2} polytypes, \textit{ex situ} control experiments, comparison of neighboring ternary oxide structures in chemical potential space, and Pareto frontier of predicted reactions forming \ce{Y2Mn2O7}
    \item  Full set of enumerated chemical reactions, experimental raw data, processed data, data processing scripts, and figure plotting scripts
\end{itemize}



\section{Acknowledgements}

This work was supported as part of GENESIS: A Next Generation Synthesis Center, an Energy Frontier Research Center funded by the U.S. Department of Energy, Office of Science, Basic Energy Sciences under Award Number DE-SC0019212. We would also like to acknowledge the facilities at 17-BM-B at the Advanced Photon Source at Argonne National Laboratory and, in particular, the support of A. Yakovenko, W. Wu. JRN acknowledges partial support from a Sloan Research Fellowship.

Theoretical calculations completed in this research used resources of the National Energy Research Scientific Computing Center (NERSC), a U.S. Department of Energy Office of Science User Facility operated under Contract No. DE-AC02-05CH11231.

\bibliography{main_r}

\providecommand{\latin}[1]{#1}
\makeatletter
\providecommand{\doi}
  {\begingroup\let\do\@makeother\dospecials
  \catcode`\{=1 \catcode`\}=2 \doi@aux}
\providecommand{\doi@aux}[1]{\endgroup\texttt{#1}}
\makeatother
\providecommand*\mcitethebibliography{\thebibliography}
\csname @ifundefined\endcsname{endmcitethebibliography}
  {\let\endmcitethebibliography\endthebibliography}{}
\begin{mcitethebibliography}{45}
\providecommand*\natexlab[1]{#1}
\providecommand*\mciteSetBstSublistMode[1]{}
\providecommand*\mciteSetBstMaxWidthForm[2]{}
\providecommand*\mciteBstWouldAddEndPuncttrue
  {\def\EndOfBibitem{\unskip.}}
\providecommand*\mciteBstWouldAddEndPunctfalse
  {\let\EndOfBibitem\relax}
\providecommand*\mciteSetBstMidEndSepPunct[3]{}
\providecommand*\mciteSetBstSublistLabelBeginEnd[3]{}
\providecommand*\EndOfBibitem{}
\mciteSetBstSublistMode{f}
\mciteSetBstMaxWidthForm{subitem}{(\alph{mcitesubitemcount})}
\mciteSetBstSublistLabelBeginEnd
  {\mcitemaxwidthsubitemform\space}
  {\relax}
  {\relax}

\bibitem[Stein \latin{et~al.}(1993)Stein, Keller, and Mallouk]{Stein1993}
Stein,~A.; Keller,~S.~W.; Mallouk,~T.~E. {Turning Down the Heat : Design and
  Mechanism in Solid-State Synthesis}. \emph{Science} \textbf{1993},
  \emph{259}, 1558--1564\relax
\mciteBstWouldAddEndPuncttrue
\mciteSetBstMidEndSepPunct{\mcitedefaultmidpunct}
{\mcitedefaultendpunct}{\mcitedefaultseppunct}\relax
\EndOfBibitem
\bibitem[Murphy \latin{et~al.}(1977)Murphy, Cros, {Di Salvo}, and
  Waszczak]{Murphy1977}
Murphy,~D.~W.; Cros,~C.; {Di Salvo},~F.~J.; Waszczak,~J.~V. {Preparation and
  properties of Li$_{x}$VS$_2$ (0 $\leq$ x $\leq$ 1)}. \emph{Inorg. Chem.}
  \textbf{1977}, \emph{16}, 3027--3031\relax
\mciteBstWouldAddEndPuncttrue
\mciteSetBstMidEndSepPunct{\mcitedefaultmidpunct}
{\mcitedefaultendpunct}{\mcitedefaultseppunct}\relax
\EndOfBibitem
\bibitem[Tsujimoto \latin{et~al.}(2007)Tsujimoto, Tassel, Hayashi, Watanabe,
  Kageyama, Yoshimura, Takano, Ceretti, Ritter, and Paulus]{Tsujimoto2007}
Tsujimoto,~Y.; Tassel,~C.; Hayashi,~N.; Watanabe,~T.; Kageyama,~H.;
  Yoshimura,~K.; Takano,~M.; Ceretti,~M.; Ritter,~C.; Paulus,~W.
  {Infinite-layer iron oxide with a square-planar coordination}. \emph{Nature}
  \textbf{2007}, \emph{450}, 1062--1065\relax
\mciteBstWouldAddEndPuncttrue
\mciteSetBstMidEndSepPunct{\mcitedefaultmidpunct}
{\mcitedefaultendpunct}{\mcitedefaultseppunct}\relax
\EndOfBibitem
\bibitem[Yajima \latin{et~al.}(2015)Yajima, Takeiri, Aidzu, Akamatsu, Fujita,
  Yoshimune, Ohkura, Lei, Gopalan, Tanaka, Brown, Green, Yamamoto, Kobayashi,
  and Kageyama]{Yajima:2015db}
Yajima,~T.; Takeiri,~F.; Aidzu,~K.; Akamatsu,~H.; Fujita,~K.; Yoshimune,~W.;
  Ohkura,~M.; Lei,~S.; Gopalan,~V.; Tanaka,~K.; Brown,~C.~M.; Green,~M.~A.;
  Yamamoto,~T.; Kobayashi,~Y.; Kageyama,~H. A labile hydride strategy for the
  synthesis of heavily nitridized BaTiO$_3$. \emph{Nature Chemistry}
  \textbf{2015}, \emph{7}, 1017--1023\relax
\mciteBstWouldAddEndPuncttrue
\mciteSetBstMidEndSepPunct{\mcitedefaultmidpunct}
{\mcitedefaultendpunct}{\mcitedefaultseppunct}\relax
\EndOfBibitem
\bibitem[Armand and Tarascon(2008)Armand, and Tarascon]{Armand2008}
Armand,~M.; Tarascon,~J.~M. Building better batteries. \emph{Nature}
  \textbf{2008}, \emph{451}, 652--657\relax
\mciteBstWouldAddEndPuncttrue
\mciteSetBstMidEndSepPunct{\mcitedefaultmidpunct}
{\mcitedefaultendpunct}{\mcitedefaultseppunct}\relax
\EndOfBibitem
\bibitem[Havelia \latin{et~al.}(2013)Havelia, Wang, Balasubramaniam, Schultz,
  Rohrer, and Salvador]{Havelia2013}
Havelia,~S.; Wang,~S.; Balasubramaniam,~K.~R.; Schultz,~A.~M.; Rohrer,~G.~S.;
  Salvador,~P.~A. Combinatorial substrate epitaxy: a new approach to growth of
  complex metastable compounds. \emph{CrystEngComm} \textbf{2013}, \emph{15},
  5434--5441\relax
\mciteBstWouldAddEndPuncttrue
\mciteSetBstMidEndSepPunct{\mcitedefaultmidpunct}
{\mcitedefaultendpunct}{\mcitedefaultseppunct}\relax
\EndOfBibitem
\bibitem[Ding \latin{et~al.}(2016)Ding, Dwaraknath, Garten, Ndione, Ginley, and
  Persson]{Ding2016}
Ding,~H.; Dwaraknath,~S.~S.; Garten,~L.; Ndione,~P.; Ginley,~D.; Persson,~K.~A.
  Computational Approach for Epitaxial Polymorph Stabilization through
  Substrate Selection. \emph{ACS Applied Materials \& Interfaces}
  \textbf{2016}, \emph{8}, 13086--13093\relax
\mciteBstWouldAddEndPuncttrue
\mciteSetBstMidEndSepPunct{\mcitedefaultmidpunct}
{\mcitedefaultendpunct}{\mcitedefaultseppunct}\relax
\EndOfBibitem
\bibitem[Miura \latin{et~al.}(2021)Miura, Bartel, Goto, Mizuguchi, Moriyoshi,
  Kuroiwa, Wang, Yaguchi, Shirai, Nagao, Rosero-Navarro, Tadanaga, Ceder, and
  Sun]{Miura2020b}
Miura,~A.; Bartel,~C.~J.; Goto,~Y.; Mizuguchi,~Y.; Moriyoshi,~C.; Kuroiwa,~Y.;
  Wang,~Y.; Yaguchi,~T.; Shirai,~M.; Nagao,~M.; Rosero-Navarro,~N.~C.;
  Tadanaga,~K.; Ceder,~G.; Sun,~W. Observing and Modeling the Sequential
  Pairwise Reactions that Drive Solid-State Ceramic Synthesis. \emph{Advanced
  Materials} \textbf{2021}, \emph{33}, 2100312\relax
\mciteBstWouldAddEndPuncttrue
\mciteSetBstMidEndSepPunct{\mcitedefaultmidpunct}
{\mcitedefaultendpunct}{\mcitedefaultseppunct}\relax
\EndOfBibitem
\bibitem[Bianchini \latin{et~al.}(2020)Bianchini, Wang, Cl{\'{e}}ment, Ouyang,
  Xiao, Kitchaev, Shi, Zhang, Wang, Kim, Zhang, Bai, Wang, Sun, and
  Ceder]{Bianchini2020}
Bianchini,~M.; Wang,~J.; Cl{\'{e}}ment,~R.~J.; Ouyang,~B.; Xiao,~P.;
  Kitchaev,~D.; Shi,~T.; Zhang,~Y.; Wang,~Y.; Kim,~H.; Zhang,~M.; Bai,~J.;
  Wang,~F.; Sun,~W.; Ceder,~G. {The interplay between thermodynamics and
  kinetics in the solid-state synthesis of layered oxides}. \emph{Nat. Mater.}
  \textbf{2020}, \emph{19}, 1088–1095\relax
\mciteBstWouldAddEndPuncttrue
\mciteSetBstMidEndSepPunct{\mcitedefaultmidpunct}
{\mcitedefaultendpunct}{\mcitedefaultseppunct}\relax
\EndOfBibitem
\bibitem[Bonneau \latin{et~al.}(1991)Bonneau, Jarvis, and Kaner]{Bonneau1991}
Bonneau,~P.~R.; Jarvis,~R.~F.; Kaner,~R.~B. {Rapid solid-state synthesis of
  materials from molybdenum disulphide to refractories}. \emph{Nature}
  \textbf{1991}, \emph{349}, 510--512\relax
\mciteBstWouldAddEndPuncttrue
\mciteSetBstMidEndSepPunct{\mcitedefaultmidpunct}
{\mcitedefaultendpunct}{\mcitedefaultseppunct}\relax
\EndOfBibitem
\bibitem[Wiley and Kaner(1992)Wiley, and Kaner]{wiley1992rapid}
Wiley,~J.~B.; Kaner,~R.~B. Rapid solid-state precursor synthesis of materials.
  \emph{Science} \textbf{1992}, \emph{255}, 1093--1097\relax
\mciteBstWouldAddEndPuncttrue
\mciteSetBstMidEndSepPunct{\mcitedefaultmidpunct}
{\mcitedefaultendpunct}{\mcitedefaultseppunct}\relax
\EndOfBibitem
\bibitem[Miura \latin{et~al.}(2020)Miura, Ito, Bartel, Sun, Rosero-Navarro,
  Tadanaga, Nakata, Maeda, and Ceder]{Miura2020a}
Miura,~A.; Ito,~H.; Bartel,~C.~J.; Sun,~W.; Rosero-Navarro,~N.~C.;
  Tadanaga,~K.; Nakata,~H.; Maeda,~K.; Ceder,~G. Selective metathesis synthesis
  of MgCr$_2$S$_4$ by control of thermodynamic driving forces. \emph{Mater.
  Horiz.} \textbf{2020}, \emph{7}, 1310--1316\relax
\mciteBstWouldAddEndPuncttrue
\mciteSetBstMidEndSepPunct{\mcitedefaultmidpunct}
{\mcitedefaultendpunct}{\mcitedefaultseppunct}\relax
\EndOfBibitem
\bibitem[Wustrow \latin{et~al.}(2018)Wustrow, Key, Phillips, Sa, Lipton, Klie,
  Vaughey, and Poeppelmeier]{Wustrow2018}
Wustrow,~A.; Key,~B.; Phillips,~P.~J.; Sa,~N.; Lipton,~A.~S.; Klie,~R.~F.;
  Vaughey,~J.~T.; Poeppelmeier,~K.~R. {Synthesis and Characterization of
  MgCr$_2$S$_4$ Thiospinel as a Potential Magnesium Cathode}. \emph{Inorg.
  Chem.} \textbf{2018}, \emph{57}, 8634--8638\relax
\mciteBstWouldAddEndPuncttrue
\mciteSetBstMidEndSepPunct{\mcitedefaultmidpunct}
{\mcitedefaultendpunct}{\mcitedefaultseppunct}\relax
\EndOfBibitem
\bibitem[Seshadri \latin{et~al.}(2012)Seshadri, Brock, Ramirez, Subramanian,
  and Thompson]{Seshadri2012}
Seshadri,~R.; Brock,~S.~L.; Ramirez,~A.; Subramanian,~M.; Thompson,~M.~E.
  {Advances in the development and growth of functional materials: Toward the
  paradigm of materials by design}. \emph{MRS Bull.} \textbf{2012}, \emph{37},
  682--690\relax
\mciteBstWouldAddEndPuncttrue
\mciteSetBstMidEndSepPunct{\mcitedefaultmidpunct}
{\mcitedefaultendpunct}{\mcitedefaultseppunct}\relax
\EndOfBibitem
\bibitem[Todd and Neilson(2019)Todd, and Neilson]{Todd2019}
Todd,~P.~K.; Neilson,~J.~R. {Selective Formation of Yttrium Manganese Oxides
  through Kinetically Competent Assisted Metathesis Reactions}. \emph{J. Am.
  Chem. Soc.} \textbf{2019}, \emph{141}, 1191--1195\relax
\mciteBstWouldAddEndPuncttrue
\mciteSetBstMidEndSepPunct{\mcitedefaultmidpunct}
{\mcitedefaultendpunct}{\mcitedefaultseppunct}\relax
\EndOfBibitem
\bibitem[Kamata \latin{et~al.}(1979)Kamata, Nakajima, and Nakamura]{Kamata1979}
Kamata,~K.; Nakajima,~T.; Nakamura,~T. {Thermogravimetric study of rare earth
  manganites AMnO$_3$ (A=Sm,Dy,Y,Er,Yb) at 1200°C}. \emph{Mater. Res. Bull.}
  \textbf{1979}, \emph{14}, 1007--1012\relax
\mciteBstWouldAddEndPuncttrue
\mciteSetBstMidEndSepPunct{\mcitedefaultmidpunct}
{\mcitedefaultendpunct}{\mcitedefaultseppunct}\relax
\EndOfBibitem
\bibitem[Balakirev and Golikov(2003)Balakirev, and Golikov]{golikov}
Balakirev,~V.~F.; Golikov,~Y.~V. Heterogeneous Phase Equilibria in Rare
  Earth--Mn--O Systems in Air. \emph{Inorganic Materials} \textbf{2003},
  \emph{39}, S1--S10\relax
\mciteBstWouldAddEndPuncttrue
\mciteSetBstMidEndSepPunct{\mcitedefaultmidpunct}
{\mcitedefaultendpunct}{\mcitedefaultseppunct}\relax
\EndOfBibitem
\bibitem[Fujinaka \latin{et~al.}(1979)Fujinaka, Kinomura, Koizumi, Miyamoto,
  and Kume]{Fujinaka1979}
Fujinaka,~H.; Kinomura,~N.; Koizumi,~M.; Miyamoto,~Y.; Kume,~S. {Syntheses and
  physical properties of pyrochlore-type A$_2$B$_2$O$_7$ (A = Tl,Y; B =
  Cr,Mn)}. \emph{Mater. Res. Bull.} \textbf{1979}, \emph{14}, 1133--1137\relax
\mciteBstWouldAddEndPuncttrue
\mciteSetBstMidEndSepPunct{\mcitedefaultmidpunct}
{\mcitedefaultendpunct}{\mcitedefaultseppunct}\relax
\EndOfBibitem
\bibitem[Subramanian \latin{et~al.}(1988)Subramanian, Torardi, Johnson,
  Pannetier, and Sleight]{Subramanian1988}
Subramanian,~M.; Torardi,~C.; Johnson,~D.; Pannetier,~J.; Sleight,~A.
  {Ferromagnetic R$_{2}$Mn$_{2}$O$_{7}$ pyrochlores (R = Dy, Lu, Y)}. \emph{J.
  Solid State Chem.} \textbf{1988}, \emph{72}, 24--30\relax
\mciteBstWouldAddEndPuncttrue
\mciteSetBstMidEndSepPunct{\mcitedefaultmidpunct}
{\mcitedefaultendpunct}{\mcitedefaultseppunct}\relax
\EndOfBibitem
\bibitem[Gardner \latin{et~al.}(2010)Gardner, Gingras, and
  Greedan]{Gardner2010a}
Gardner,~J.~S.; Gingras,~M. J.~P.; Greedan,~J.~E. {Magnetic pyrochlore oxides}.
  \emph{Rev. Mod. Phys.} \textbf{2010}, \emph{82}, 53--107\relax
\mciteBstWouldAddEndPuncttrue
\mciteSetBstMidEndSepPunct{\mcitedefaultmidpunct}
{\mcitedefaultendpunct}{\mcitedefaultseppunct}\relax
\EndOfBibitem
\bibitem[Todd \latin{et~al.}(2019)Todd, Smith, and Neilson]{Todd2019a}
Todd,~P.~K.; Smith,~A. M.~M.; Neilson,~J.~R. {Yttrium Manganese Oxide Phase
  Stability and Selectivity Using Lithium Carbonate Assisted Metathesis
  Reactions}. \emph{Inorg. Chem.} \textbf{2019}, \emph{58}, 15166--15174\relax
\mciteBstWouldAddEndPuncttrue
\mciteSetBstMidEndSepPunct{\mcitedefaultmidpunct}
{\mcitedefaultendpunct}{\mcitedefaultseppunct}\relax
\EndOfBibitem
\bibitem[Todd \latin{et~al.}(2020)Todd, Wustrow, McAuliffe, McDermott, Tran,
  McBride, Boeding, O'Nolan, Liu, Dwaraknath, Chapman, Billinge, Persson, Huq,
  Veith, and Neilson]{Todd2020}
Todd,~P.~K. \latin{et~al.}  Defect-Accommodating Intermediates Yield Selective
  Low-Temperature Synthesis of YMnO$_3$ Polymorphs. \emph{Inorg. Chem.}
  \textbf{2020}, \emph{59}, 13639--13650\relax
\mciteBstWouldAddEndPuncttrue
\mciteSetBstMidEndSepPunct{\mcitedefaultmidpunct}
{\mcitedefaultendpunct}{\mcitedefaultseppunct}\relax
\EndOfBibitem
\bibitem[Chupas \latin{et~al.}(2008)Chupas, Chapman, Kurtz, Hanson, Lee, and
  Grey]{Chupas2008}
Chupas,~P.~J.; Chapman,~K.~W.; Kurtz,~C.; Hanson,~J.~C.; Lee,~P.~L.;
  Grey,~C.~P. {A versatile sample-environment cell for non-ambient X-ray
  scattering experiments}. \emph{J. Appl. Crystallogr.} \textbf{2008},
  \emph{41}, 822--824\relax
\mciteBstWouldAddEndPuncttrue
\mciteSetBstMidEndSepPunct{\mcitedefaultmidpunct}
{\mcitedefaultendpunct}{\mcitedefaultseppunct}\relax
\EndOfBibitem
\bibitem[Jiang \latin{et~al.}(2017)Jiang, Ramanathan, and Shoemaker]{Jiang2016}
Jiang,~Z.; Ramanathan,~A.; Shoemaker,~D.~P. {In situ identification of kinetic
  factors that expedite inorganic crystal formation and discovery}. \emph{J.
  Mater. Chem. C} \textbf{2017}, \emph{5}, 5709--5717\relax
\mciteBstWouldAddEndPuncttrue
\mciteSetBstMidEndSepPunct{\mcitedefaultmidpunct}
{\mcitedefaultendpunct}{\mcitedefaultseppunct}\relax
\EndOfBibitem
\bibitem[Yin \latin{et~al.}(2018)Yin, Mattei, Li, Zheng, Zhao, Omenya, Fang,
  Li, Li, Xie, Zhang, Whittingham, Meng, Manthiram, and Khalifah]{Yin2018}
Yin,~L.; Mattei,~G.~S.; Li,~Z.; Zheng,~J.; Zhao,~W.; Omenya,~F.; Fang,~C.;
  Li,~W.; Li,~J.; Xie,~Q.; Zhang,~J.-G.; Whittingham,~M.~S.; Meng,~Y.~S.;
  Manthiram,~A.; Khalifah,~P.~G. Extending the limits of powder diffraction
  analysis: Diffraction parameter space, occupancy defects, and atomic form
  factors. \emph{Review of Scientific Instruments} \textbf{2018}, \emph{89},
  093002\relax
\mciteBstWouldAddEndPuncttrue
\mciteSetBstMidEndSepPunct{\mcitedefaultmidpunct}
{\mcitedefaultendpunct}{\mcitedefaultseppunct}\relax
\EndOfBibitem
\bibitem[Yin \latin{et~al.}(2020)Yin, Li, Mattei, Zheng, Zhao, Omenya, Fang,
  Li, Li, Xie, Erickson, Zhang, Whittingham, Meng, Manthiram, and
  Khalifah]{Yin2020}
Yin,~L. \latin{et~al.}  {Thermodynamics of Antisite Defects in Layered NMC
  Cathodes: Systematic Insights from High-Precision Powder Diffraction
  Analyses}. \emph{Chem. Mater.} \textbf{2020}, \emph{32}, 1002--1010\relax
\mciteBstWouldAddEndPuncttrue
\mciteSetBstMidEndSepPunct{\mcitedefaultmidpunct}
{\mcitedefaultendpunct}{\mcitedefaultseppunct}\relax
\EndOfBibitem
\bibitem[Jain \latin{et~al.}(2013)Jain, Ong, Hautier, Chen, Richards, Dacek,
  Cholia, Gunter, Skinner, Ceder, and Persson]{Jain2013}
Jain,~A.; Ong,~S.~P.; Hautier,~G.; Chen,~W.; Richards,~W.~D.; Dacek,~S.;
  Cholia,~S.; Gunter,~D.; Skinner,~D.; Ceder,~G.; Persson,~K.~A. {Commentary:
  The Materials Project: A materials genome approach to accelerating materials
  innovation}. \emph{APL Mater.} \textbf{2013}, \emph{1}, 011002\relax
\mciteBstWouldAddEndPuncttrue
\mciteSetBstMidEndSepPunct{\mcitedefaultmidpunct}
{\mcitedefaultendpunct}{\mcitedefaultseppunct}\relax
\EndOfBibitem
\bibitem[Bartel \latin{et~al.}(2018)Bartel, Millican, Deml, Rumptz, Tumas,
  Weimer, Lany, Stevanović, Musgrave, and Holder]{Bartel2018}
Bartel,~C.~J.; Millican,~S.~L.; Deml,~A.~M.; Rumptz,~J.~R.; Tumas,~W.;
  Weimer,~A.~W.; Lany,~S.; Stevanović,~V.; Musgrave,~C.~B.; Holder,~A.~M.
  Physical descriptor for the {Gibbs} energy of inorganic crystalline solids
  and temperature-dependent materials chemistry. \emph{Nature Communications}
  \textbf{2018}, \emph{9}, 4168\relax
\mciteBstWouldAddEndPuncttrue
\mciteSetBstMidEndSepPunct{\mcitedefaultmidpunct}
{\mcitedefaultendpunct}{\mcitedefaultseppunct}\relax
\EndOfBibitem
\bibitem[Ong \latin{et~al.}(2010)Ong, Jain, Hautier, Kang, and Ceder]{Ong2010}
Ong,~S.~P.; Jain,~A.; Hautier,~G.; Kang,~B.; Ceder,~G. {Thermal stabilities of
  delithiated olivine MPO$_4$ (M=Fe, Mn) cathodes investigated using first
  principles calculations}. \emph{Electrochem. Commun.} \textbf{2010},
  \emph{12}, 427--430\relax
\mciteBstWouldAddEndPuncttrue
\mciteSetBstMidEndSepPunct{\mcitedefaultmidpunct}
{\mcitedefaultendpunct}{\mcitedefaultseppunct}\relax
\EndOfBibitem
\bibitem[Yokokawa(1999)]{Yokokawa1999}
Yokokawa,~H. {Generalized chemical potential diagram and its applications to
  chemical reactions at interfaces between dissimilar materials}. \emph{Journal
  of Phase Equilibria} \textbf{1999}, \emph{20}, 258\relax
\mciteBstWouldAddEndPuncttrue
\mciteSetBstMidEndSepPunct{\mcitedefaultmidpunct}
{\mcitedefaultendpunct}{\mcitedefaultseppunct}\relax
\EndOfBibitem
\bibitem[Patel \latin{et~al.}(2019)Patel, Nørskov, Persson, and
  Montoya]{Montoya2019}
Patel,~A.~M.; Nørskov,~J.~K.; Persson,~K.~A.; Montoya,~J.~H. Efficient
  Pourbaix diagrams of many-element compounds. \emph{Phys. Chem. Chem. Phys.}
  \textbf{2019}, \emph{21}, 25323--25327\relax
\mciteBstWouldAddEndPuncttrue
\mciteSetBstMidEndSepPunct{\mcitedefaultmidpunct}
{\mcitedefaultendpunct}{\mcitedefaultseppunct}\relax
\EndOfBibitem
\bibitem[Virtanen \latin{et~al.}(2020)Virtanen, Gommers, Oliphant, Haberland,
  Reddy, Cournapeau, Burovski, Peterson, Weckesser, Bright, van~der Walt,
  Brett, Wilson, Millman, Mayorov, Nelson, Jones, Kern, Larson, Carey, Polat,
  Feng, Moore, VanderPlas, Laxalde, Perktold, Cimrman, Henriksen, Quintero,
  Harris, Archibald, Ribeiro, Pedregosa, van Mulbregt, Vijaykumar, Bardelli,
  Rothberg, Hilboll, Kloeckner, Scopatz, Lee, Rokem, Woods, Fulton, Masson,
  H{\"{a}}ggstr{\"{o}}m, Fitzgerald, Nicholson, Hagen, Pasechnik, Olivetti,
  Martin, Wieser, Silva, Lenders, Wilhelm, Young, Price, Ingold, Allen, Lee,
  Audren, Probst, Dietrich, Silterra, Webber, Slavi{\v{c}}, Nothman, Buchner,
  Kulick, Sch{\"{o}}nberger, {de Miranda Cardoso}, Reimer, Harrington,
  Rodr{\'{i}}guez, Nunez-Iglesias, Kuczynski, Tritz, Thoma, Newville,
  K{\"{u}}mmerer, Bolingbroke, Tartre, Pak, Smith, Nowaczyk, Shebanov, Pavlyk,
  Brodtkorb, Lee, McGibbon, Feldbauer, Lewis, Tygier, Sievert, Vigna, Peterson,
  More, Pudlik, Oshima, Pingel, Robitaille, Spura, Jones, Cera, Leslie, Zito,
  Krauss, Upadhyay, Halchenko, V{\'{a}}zquez-Baeza, and
  Contributors]{Virtanen2020}
Virtanen,~P. \latin{et~al.}  {SciPy 1.0: fundamental algorithms for scientific
  computing in Python}. \emph{Nature Methods} \textbf{2020}, \emph{17},
  261--272\relax
\mciteBstWouldAddEndPuncttrue
\mciteSetBstMidEndSepPunct{\mcitedefaultmidpunct}
{\mcitedefaultendpunct}{\mcitedefaultseppunct}\relax
\EndOfBibitem
\bibitem[McDermott \latin{et~al.}(2021)McDermott, Dwaraknath, and
  Persson]{McDermott2021}
McDermott,~M.~J.; Dwaraknath,~S.~S.; Persson,~K.~A. A graph-based network for
  predicting chemical reaction pathways in solid-state materials synthesis.
  \emph{Nature Communications} \textbf{2021}, \emph{12}, 3097\relax
\mciteBstWouldAddEndPuncttrue
\mciteSetBstMidEndSepPunct{\mcitedefaultmidpunct}
{\mcitedefaultendpunct}{\mcitedefaultseppunct}\relax
\EndOfBibitem
\bibitem[McDermott and Dwaraknath(2021)McDermott, and Dwaraknath]{zenodo}
McDermott,~M.; Dwaraknath,~S. GENESIS-EFRC/reaction-network: v2.0.3. 2021;
  \url{https://doi.org/10.5281/zenodo.5165276}\relax
\mciteBstWouldAddEndPuncttrue
\mciteSetBstMidEndSepPunct{\mcitedefaultmidpunct}
{\mcitedefaultendpunct}{\mcitedefaultseppunct}\relax
\EndOfBibitem
\bibitem[Caballero \latin{et~al.}(2002)Caballero, Hern{\'{a}}n, Morales,
  S{\'{a}}nchez, {Santos Pe{\~{n}}a}, and Aranda]{Caballero2002}
Caballero,~A.; Hern{\'{a}}n,~L.; Morales,~J.; S{\'{a}}nchez,~L.; {Santos
  Pe{\~{n}}a},~J.; Aranda,~M.~A. {Synthesis and characterization of
  high-temperature hexagonal P2-Na$_{0.6}$MnO$_2$ and its electrochemical
  behaviour as cathode in sodium cells}. \emph{J. Mater. Chem.} \textbf{2002},
  \emph{12}, 1142--1147\relax
\mciteBstWouldAddEndPuncttrue
\mciteSetBstMidEndSepPunct{\mcitedefaultmidpunct}
{\mcitedefaultendpunct}{\mcitedefaultseppunct}\relax
\EndOfBibitem
\bibitem[Paulsen and Dahn(1999)Paulsen, and Dahn]{Co1999}
Paulsen,~J.; Dahn,~J. {Studies of the layered manganese bronzes,
  Na$_{2/3}$[Mn$_{1-x}$M$_{x}$]O$_{2}$ with M=Co, Ni, Li, and
  Li$_{2/3}$[Mn$_{1-x}$M$_{x}$]O$_{2}$ prepared by ion-exchange}. \emph{Solid
  State Ionics} \textbf{1999}, \emph{126}, 3--24\relax
\mciteBstWouldAddEndPuncttrue
\mciteSetBstMidEndSepPunct{\mcitedefaultmidpunct}
{\mcitedefaultendpunct}{\mcitedefaultseppunct}\relax
\EndOfBibitem
\bibitem[Cl{\'{e}}ment \latin{et~al.}(2015)Cl{\'{e}}ment, Bruce, and
  Grey]{Clement2015}
Cl{\'{e}}ment,~R.~J.; Bruce,~P.~G.; Grey,~C.~P. {Review-Manganese-based P2-type
  transition metal oxides as sodium-ion battery cathode materials}. \emph{J.
  Electrochem. Soc.} \textbf{2015}, \emph{162}, A2589--A2604\relax
\mciteBstWouldAddEndPuncttrue
\mciteSetBstMidEndSepPunct{\mcitedefaultmidpunct}
{\mcitedefaultendpunct}{\mcitedefaultseppunct}\relax
\EndOfBibitem
\bibitem[Jain \latin{et~al.}(2011)Jain, Hautier, {Ping Ong}, Moore, Fischer,
  Persson, and Ceder]{Jain2011}
Jain,~A.; Hautier,~G.; {Ping Ong},~S.; Moore,~C.~J.; Fischer,~C.~C.;
  Persson,~K.~A.; Ceder,~G. {Formation enthalpies by mixing GGA and GGA + U
  calculations}. \emph{Physical Review B} \textbf{2011}, \emph{84}, 45115\relax
\mciteBstWouldAddEndPuncttrue
\mciteSetBstMidEndSepPunct{\mcitedefaultmidpunct}
{\mcitedefaultendpunct}{\mcitedefaultseppunct}\relax
\EndOfBibitem
\bibitem[Martinolich \latin{et~al.}(2016)Martinolich, Higgins, Shores, and
  Neilson]{Martinolich2016b}
Martinolich,~A.~J.; Higgins,~R.~F.; Shores,~M.~P.; Neilson,~J.~R. {Lewis Base
  Mediated Polymorph Selectivity of Pyrite CuSe$_2$ through Atom Transfer in
  Solid-State Metathesis}. \emph{Chem. Mater.} \textbf{2016}, \emph{28},
  1854--1860\relax
\mciteBstWouldAddEndPuncttrue
\mciteSetBstMidEndSepPunct{\mcitedefaultmidpunct}
{\mcitedefaultendpunct}{\mcitedefaultseppunct}\relax
\EndOfBibitem
\bibitem[Schmalzried(1981)]{Schmalzried1981}
Schmalzried,~H. \emph{Solid state reactions}; Verlag Chemie, 1981\relax
\mciteBstWouldAddEndPuncttrue
\mciteSetBstMidEndSepPunct{\mcitedefaultmidpunct}
{\mcitedefaultendpunct}{\mcitedefaultseppunct}\relax
\EndOfBibitem
\bibitem[Kitchaev \latin{et~al.}(2017)Kitchaev, Dacek, Sun, and
  Ceder]{Kitchaev2017}
Kitchaev,~D.~A.; Dacek,~S.~T.; Sun,~W.; Ceder,~G. {Thermodynamics of Phase
  Selection in MnO$_2$ Framework Structures through Alkali Intercalation and
  Hydration}. \emph{J. Am. Chem. Soc.} \textbf{2017}, \emph{139},
  2672--2681\relax
\mciteBstWouldAddEndPuncttrue
\mciteSetBstMidEndSepPunct{\mcitedefaultmidpunct}
{\mcitedefaultendpunct}{\mcitedefaultseppunct}\relax
\EndOfBibitem
\bibitem[Bai \latin{et~al.}(2020)Bai, Sun, Zhao, Wang, Xiao, Ko, Huq, Ceder,
  and Wang]{Bai2020}
Bai,~J.; Sun,~W.; Zhao,~J.; Wang,~D.; Xiao,~P.; Ko,~J. Y.~P.; Huq,~A.;
  Ceder,~G.; Wang,~F. Kinetic {Pathways} {Templated} by {Low}-{Temperature}
  {Intermediates} during {Solid}-{State} {Synthesis} of {Layered} {Oxides}.
  \emph{Chemistry of Materials} \textbf{2020}, \emph{32}, 9906--9913,
  Publisher: American Chemical Society\relax
\mciteBstWouldAddEndPuncttrue
\mciteSetBstMidEndSepPunct{\mcitedefaultmidpunct}
{\mcitedefaultendpunct}{\mcitedefaultseppunct}\relax
\EndOfBibitem
\bibitem[Neilson and McQueen(2012)Neilson, and McQueen]{Si2012}
Neilson,~J.~R.; McQueen,~T.~M. {Bonding, Ion Mobility, and Rate-Limiting Steps
  in Deintercalation Reactions with ThCr$_2$Si$_2$-type KNi$_2$Se$_2$}.
  \emph{J. Am. Chem. Soc.} \textbf{2012}, \emph{134}, 7750--7757\relax
\mciteBstWouldAddEndPuncttrue
\mciteSetBstMidEndSepPunct{\mcitedefaultmidpunct}
{\mcitedefaultendpunct}{\mcitedefaultseppunct}\relax
\EndOfBibitem
\bibitem[Hayward \latin{et~al.}(1999)Hayward, Green, Rosseinsky, and
  Sloan]{Hayward1999}
Hayward,~M.~A.; Green,~M.~A.; Rosseinsky,~M.~J.; Sloan,~J. {Sodium hydride as a
  powerful reducing agent for topotactic oxide deintercalation: Synthesis and
  characterization of the nickel(I) oxide LaNiO$_2$}. \emph{J. Am. Chem. Soc.}
  \textbf{1999}, \emph{121}, 8843--8854\relax
\mciteBstWouldAddEndPuncttrue
\mciteSetBstMidEndSepPunct{\mcitedefaultmidpunct}
{\mcitedefaultendpunct}{\mcitedefaultseppunct}\relax
\EndOfBibitem
\end{mcitethebibliography}


\providecommand{\latin}[1]{#1}
\makeatletter
\providecommand{\doi}
  {\begingroup\let\do\@makeother\dospecials
  \catcode`\{=1 \catcode`\}=2 \doi@aux}
\providecommand{\doi@aux}[1]{\endgroup\texttt{#1}}
\makeatother
\providecommand*\mcitethebibliography{\thebibliography}
\csname @ifundefined\endcsname{endmcitethebibliography}
  {\let\endmcitethebibliography\endthebibliography}{}
\begin{mcitethebibliography}{17}
\providecommand*\natexlab[1]{#1}
\providecommand*\mciteSetBstSublistMode[1]{}
\providecommand*\mciteSetBstMaxWidthForm[2]{}
\providecommand*\mciteBstWouldAddEndPuncttrue
  {\def\EndOfBibitem{\unskip.}}
\providecommand*\mciteBstWouldAddEndPunctfalse
  {\let\EndOfBibitem\relax}
\providecommand*\mciteSetBstMidEndSepPunct[3]{}
\providecommand*\mciteSetBstSublistLabelBeginEnd[3]{}
\providecommand*\EndOfBibitem{}
\mciteSetBstSublistMode{f}
\mciteSetBstMaxWidthForm{subitem}{(\alph{mcitesubitemcount})}
\mciteSetBstSublistLabelBeginEnd
  {\mcitemaxwidthsubitemform\space}
  {\relax}
  {\relax}

\bibitem[Bartel \latin{et~al.}(2018)Bartel, Millican, Deml, Rumptz, Tumas,
  Weimer, Lany, Stevanović, Musgrave, and Holder]{Bartel2018}
Bartel,~C.~J.; Millican,~S.~L.; Deml,~A.~M.; Rumptz,~J.~R.; Tumas,~W.;
  Weimer,~A.~W.; Lany,~S.; Stevanović,~V.; Musgrave,~C.~B.; Holder,~A.~M.
  Physical descriptor for the {Gibbs} energy of inorganic crystalline solids
  and temperature-dependent materials chemistry. \emph{Nature Communications}
  \textbf{2018}, \emph{9}, 4168\relax
\mciteBstWouldAddEndPuncttrue
\mciteSetBstMidEndSepPunct{\mcitedefaultmidpunct}
{\mcitedefaultendpunct}{\mcitedefaultseppunct}\relax
\EndOfBibitem
\bibitem[Barin(2008)]{Barin_Li}
Barin,~I. \emph{Thermochemical Data of Pure Substances}; John Wiley and Sons,
  Ltd, 2008; Chapter 12, pp 925--992\relax
\mciteBstWouldAddEndPuncttrue
\mciteSetBstMidEndSepPunct{\mcitedefaultmidpunct}
{\mcitedefaultendpunct}{\mcitedefaultseppunct}\relax
\EndOfBibitem
\bibitem[Barin(2008)]{Barin_Na}
Barin,~I. \emph{Thermochemical Data of Pure Substances}; John Wiley and Sons,
  Ltd, 2008; Chapter 12, pp 1080--1237\relax
\mciteBstWouldAddEndPuncttrue
\mciteSetBstMidEndSepPunct{\mcitedefaultmidpunct}
{\mcitedefaultendpunct}{\mcitedefaultseppunct}\relax
\EndOfBibitem
\bibitem[Barin(2008)]{Barin_K}
Barin,~I. \emph{Thermochemical Data of Pure Substances}; John Wiley and Sons,
  Ltd, 2008; Chapter 12, pp 844--924\relax
\mciteBstWouldAddEndPuncttrue
\mciteSetBstMidEndSepPunct{\mcitedefaultmidpunct}
{\mcitedefaultendpunct}{\mcitedefaultseppunct}\relax
\EndOfBibitem
\bibitem[{Malcolm W. Chase}(1998)]{MalcolmW.Chase1998}
{Malcolm W. Chase},~J. \emph{{NIST-JANAF thermochemical tables}}; Fourth
  edition. Washington, DC: American Chemical Society; New York: American
  Institute of Physics for the National Institute of Standards and Technology,
  1998\relax
\mciteBstWouldAddEndPuncttrue
\mciteSetBstMidEndSepPunct{\mcitedefaultmidpunct}
{\mcitedefaultendpunct}{\mcitedefaultseppunct}\relax
\EndOfBibitem
\bibitem[Todd \latin{et~al.}(2019)Todd, Smith, and Neilson]{Todd2019a}
Todd,~P.~K.; Smith,~A. M.~M.; Neilson,~J.~R. {Yttrium Manganese Oxide Phase
  Stability and Selectivity Using Lithium Carbonate Assisted Metathesis
  Reactions}. \emph{Inorg. Chem.} \textbf{2019}, \emph{58}, 15166--15174\relax
\mciteBstWouldAddEndPuncttrue
\mciteSetBstMidEndSepPunct{\mcitedefaultmidpunct}
{\mcitedefaultendpunct}{\mcitedefaultseppunct}\relax
\EndOfBibitem
\bibitem[Bianchini \latin{et~al.}(2020)Bianchini, Wang, Cl{\'{e}}ment, Ouyang,
  Xiao, Kitchaev, Shi, Zhang, Wang, Kim, Zhang, Bai, Wang, Sun, and
  Ceder]{Bianchini2020}
Bianchini,~M.; Wang,~J.; Cl{\'{e}}ment,~R.~J.; Ouyang,~B.; Xiao,~P.;
  Kitchaev,~D.; Shi,~T.; Zhang,~Y.; Wang,~Y.; Kim,~H.; Zhang,~M.; Bai,~J.;
  Wang,~F.; Sun,~W.; Ceder,~G. {The interplay between thermodynamics and
  kinetics in the solid-state synthesis of layered oxides}. \emph{Nat. Mater.}
  \textbf{2020}, \emph{19}, 1088–1095\relax
\mciteBstWouldAddEndPuncttrue
\mciteSetBstMidEndSepPunct{\mcitedefaultmidpunct}
{\mcitedefaultendpunct}{\mcitedefaultseppunct}\relax
\EndOfBibitem
\bibitem[Takada \latin{et~al.}(2005)Takada, Osada, Izumi, Sakurai,
  Takayama-Muromachi, and Sasaki]{Takada2005}
Takada,~K.; Osada,~M.; Izumi,~F.; Sakurai,~H.; Takayama-Muromachi,~E.;
  Sasaki,~T. Characterization of Sodium Cobalt Oxides Related to P3-Phase
  Superconductor. \emph{Chemistry of Materials} \textbf{2005}, \emph{17},
  2034--2040\relax
\mciteBstWouldAddEndPuncttrue
\mciteSetBstMidEndSepPunct{\mcitedefaultmidpunct}
{\mcitedefaultendpunct}{\mcitedefaultseppunct}\relax
\EndOfBibitem
\bibitem[Caballero \latin{et~al.}(2002)Caballero, Hern{\'{a}}n, Morales,
  S{\'{a}}nchez, {Santos Pe{\~{n}}a}, and Aranda]{Caballero2002}
Caballero,~A.; Hern{\'{a}}n,~L.; Morales,~J.; S{\'{a}}nchez,~L.; {Santos
  Pe{\~{n}}a},~J.; Aranda,~M.~A. {Synthesis and characterization of
  high-temperature hexagonal P2-Na$_{0.6}$MnO$_2$ and its electrochemical
  behaviour as cathode in sodium cells}. \emph{J. Mater. Chem.} \textbf{2002},
  \emph{12}, 1142--1147\relax
\mciteBstWouldAddEndPuncttrue
\mciteSetBstMidEndSepPunct{\mcitedefaultmidpunct}
{\mcitedefaultendpunct}{\mcitedefaultseppunct}\relax
\EndOfBibitem
\bibitem[Paulsen and Dahn(1999)Paulsen, and Dahn]{Co1999}
Paulsen,~J.; Dahn,~J. {Studies of the layered manganese bronzes,
  Na$_{2/3}$[Mn$_{1-x}$M$_{x}$]O$_{2}$ with M=Co, Ni, Li, and
  Li$_{2/3}$[Mn$_{1-x}$M$_{x}$]O$_{2}$ prepared by ion-exchange}. \emph{Solid
  State Ionics} \textbf{1999}, \emph{126}, 3--24\relax
\mciteBstWouldAddEndPuncttrue
\mciteSetBstMidEndSepPunct{\mcitedefaultmidpunct}
{\mcitedefaultendpunct}{\mcitedefaultseppunct}\relax
\EndOfBibitem
\bibitem[Cl{\'{e}}ment \latin{et~al.}(2015)Cl{\'{e}}ment, Bruce, and
  Grey]{Clement2015}
Cl{\'{e}}ment,~R.~J.; Bruce,~P.~G.; Grey,~C.~P. {Review-Manganese-based P2-type
  transition metal oxides as sodium-ion battery cathode materials}. \emph{J.
  Electrochem. Soc.} \textbf{2015}, \emph{162}, A2589--A2604\relax
\mciteBstWouldAddEndPuncttrue
\mciteSetBstMidEndSepPunct{\mcitedefaultmidpunct}
{\mcitedefaultendpunct}{\mcitedefaultseppunct}\relax
\EndOfBibitem
\bibitem[Yin \latin{et~al.}(2018)Yin, Mattei, Li, Zheng, Zhao, Omenya, Fang,
  Li, Li, Xie, Zhang, Whittingham, Meng, Manthiram, and Khalifah]{Yin2018}
Yin,~L.; Mattei,~G.~S.; Li,~Z.; Zheng,~J.; Zhao,~W.; Omenya,~F.; Fang,~C.;
  Li,~W.; Li,~J.; Xie,~Q.; Zhang,~J.-G.; Whittingham,~M.~S.; Meng,~Y.~S.;
  Manthiram,~A.; Khalifah,~P.~G. Extending the limits of powder diffraction
  analysis: Diffraction parameter space, occupancy defects, and atomic form
  factors. \emph{Review of Scientific Instruments} \textbf{2018}, \emph{89},
  093002\relax
\mciteBstWouldAddEndPuncttrue
\mciteSetBstMidEndSepPunct{\mcitedefaultmidpunct}
{\mcitedefaultendpunct}{\mcitedefaultseppunct}\relax
\EndOfBibitem
\bibitem[Todd and Neilson(2019)Todd, and Neilson]{Todd2019}
Todd,~P.~K.; Neilson,~J.~R. {Selective Formation of Yttrium Manganese Oxides
  through Kinetically Competent Assisted Metathesis Reactions}. \emph{J. Am.
  Chem. Soc.} \textbf{2019}, \emph{141}, 1191--1195\relax
\mciteBstWouldAddEndPuncttrue
\mciteSetBstMidEndSepPunct{\mcitedefaultmidpunct}
{\mcitedefaultendpunct}{\mcitedefaultseppunct}\relax
\EndOfBibitem
\bibitem[Stoyanova \latin{et~al.}(2010)Stoyanova, Carlier, Sendova-Vassileva,
  Yoncheva, Zhecheva, Nihtianova, and Delmas]{Stoyanova2010}
Stoyanova,~R.; Carlier,~D.; Sendova-Vassileva,~M.; Yoncheva,~M.; Zhecheva,~E.;
  Nihtianova,~D.; Delmas,~C. {Stabilization of over-stoichiometric Mn$^{4+}$ in
  layered Na$_{2/3}$MnO$_{2}$}. \emph{J. Solid State Chem.} \textbf{2010},
  \emph{183}, 1372--1379\relax
\mciteBstWouldAddEndPuncttrue
\mciteSetBstMidEndSepPunct{\mcitedefaultmidpunct}
{\mcitedefaultendpunct}{\mcitedefaultseppunct}\relax
\EndOfBibitem
\bibitem[Kitchaev \latin{et~al.}(2017)Kitchaev, Dacek, Sun, and
  Ceder]{Kitchaev2017}
Kitchaev,~D.~A.; Dacek,~S.~T.; Sun,~W.; Ceder,~G. {Thermodynamics of Phase
  Selection in MnO$_2$ Framework Structures through Alkali Intercalation and
  Hydration}. \emph{J. Am. Chem. Soc.} \textbf{2017}, \emph{139},
  2672--2681\relax
\mciteBstWouldAddEndPuncttrue
\mciteSetBstMidEndSepPunct{\mcitedefaultmidpunct}
{\mcitedefaultendpunct}{\mcitedefaultseppunct}\relax
\EndOfBibitem
\bibitem[Aykol \latin{et~al.}(2021)Aykol, Montoya, and Hummelshøj]{Aykol2021a}
Aykol,~M.; Montoya,~J.~H.; Hummelshøj,~J. Rational {Solid}-{State} {Synthesis}
  {Routes} for {Inorganic} {Materials}. \emph{Journal of the American Chemical
  Society} \textbf{2021}, \emph{143}, 9244--9259, Publisher: American Chemical
  Society\relax
\mciteBstWouldAddEndPuncttrue
\mciteSetBstMidEndSepPunct{\mcitedefaultmidpunct}
{\mcitedefaultendpunct}{\mcitedefaultseppunct}\relax
\EndOfBibitem
\end{mcitethebibliography}

\begin{tocentry}
\begin{center}
\includegraphics[]{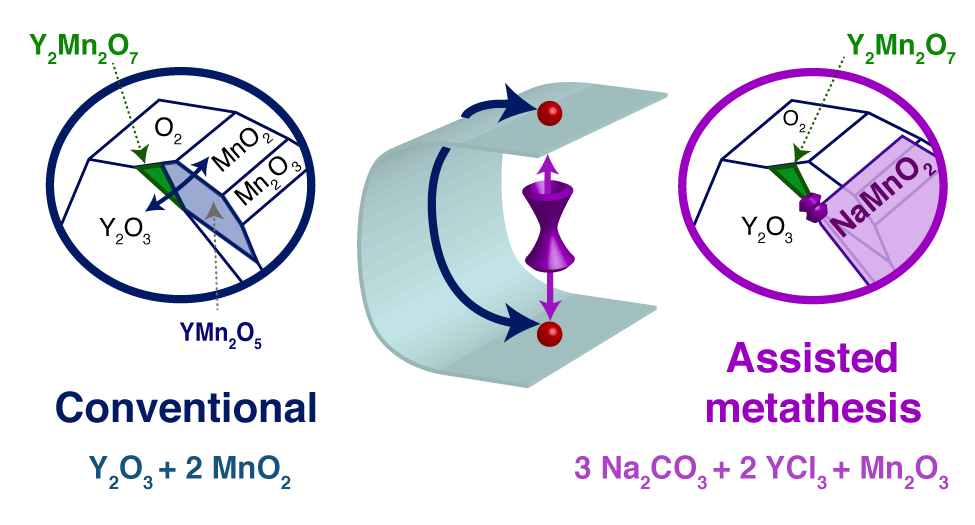}
\end{center}
\end{tocentry}

\end{document}


\newpage

\tableofcontents
\section{Assisted metathesis reaction energetics}

\begin{figure}[h]
\begin{center}
\includegraphics[width=5in]{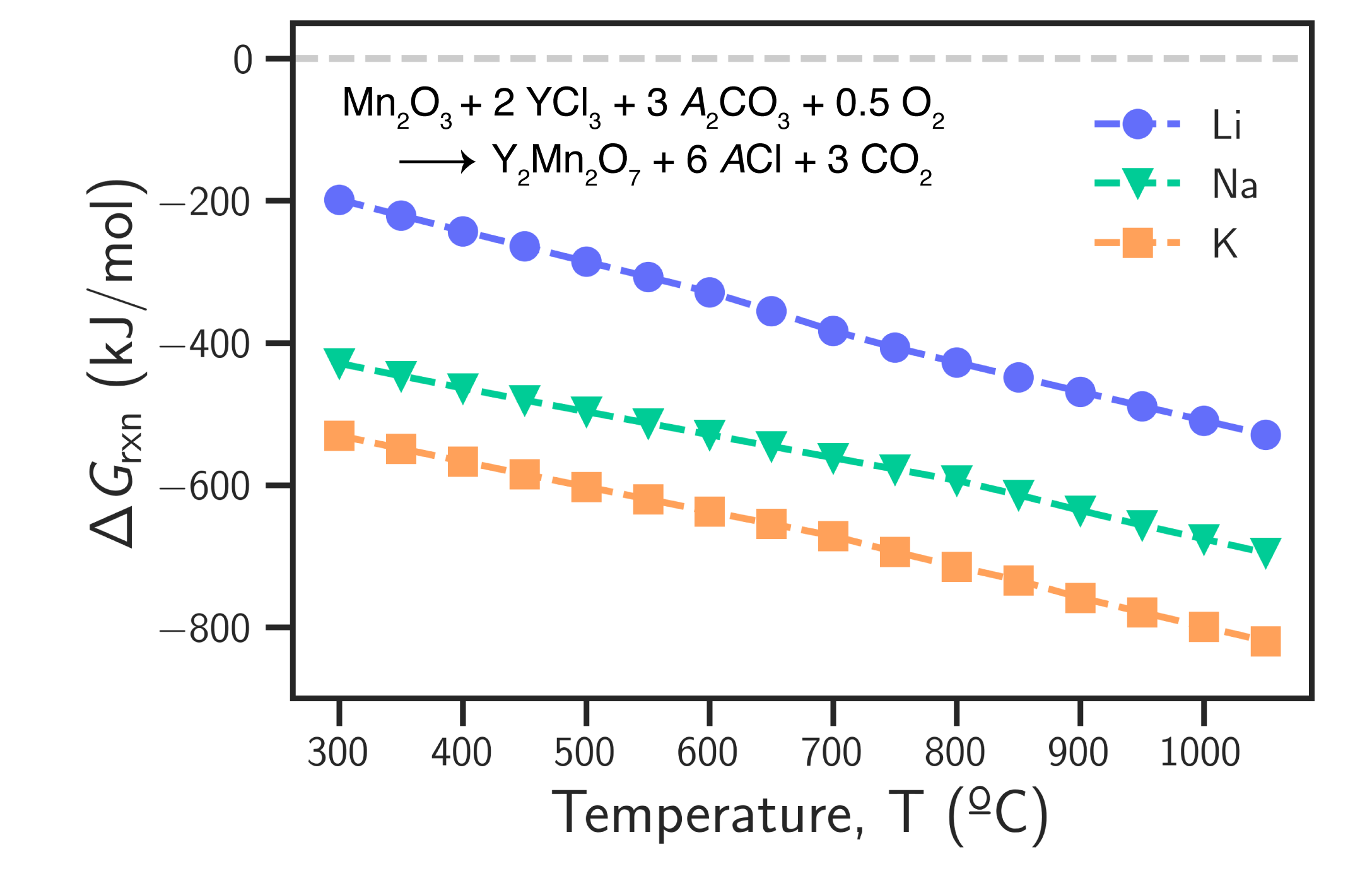}
\caption{\label{S-fig:rxn_energies} Reaction energetics of the assisted metathesis reaction, \ce{Mn2O3 + 2YCl3 + 3A2CO3 + 0.5O2 -> Y2Mn2O7 + 6ACl + 3CO2} for alkali metals A=Li, Na, and K as predicted by Materials Project data and machine-learned descriptor for solids. \cite{Bartel2018} Thermochemical data for \ce{A2CO3} and \ce{ACl} were acquired from experimental tables \cite{Barin_Li, Barin_Na, Barin_K} and \ce{CO2} from the NIST-JANAF database\cite{MalcolmW.Chase1998};  hence changes in slope are due to the melting of \ce{A2CO3} and/or \ce{ACl}.}
\end{center} 
\end{figure}

The assisted metathesis net reaction \ce{Mn2O3 + 2YCl3 + 3A2CO3 + 0.5O2 -> Y2Mn2O7 + 6ACl + 3CO2} (A= Li, Na, K) is predicted to be exergonic ($G < 0$) for all temperatures (Figure~\ref{S-fig:rxn_energies}). The least negative reaction energy occurs for lithium, followed by sodium, and then potassium. 

\section{Detailed description of the Na-based reaction progression}

\begin{figure}[h]
\begin{center}
\includegraphics[width=0.95\textwidth]{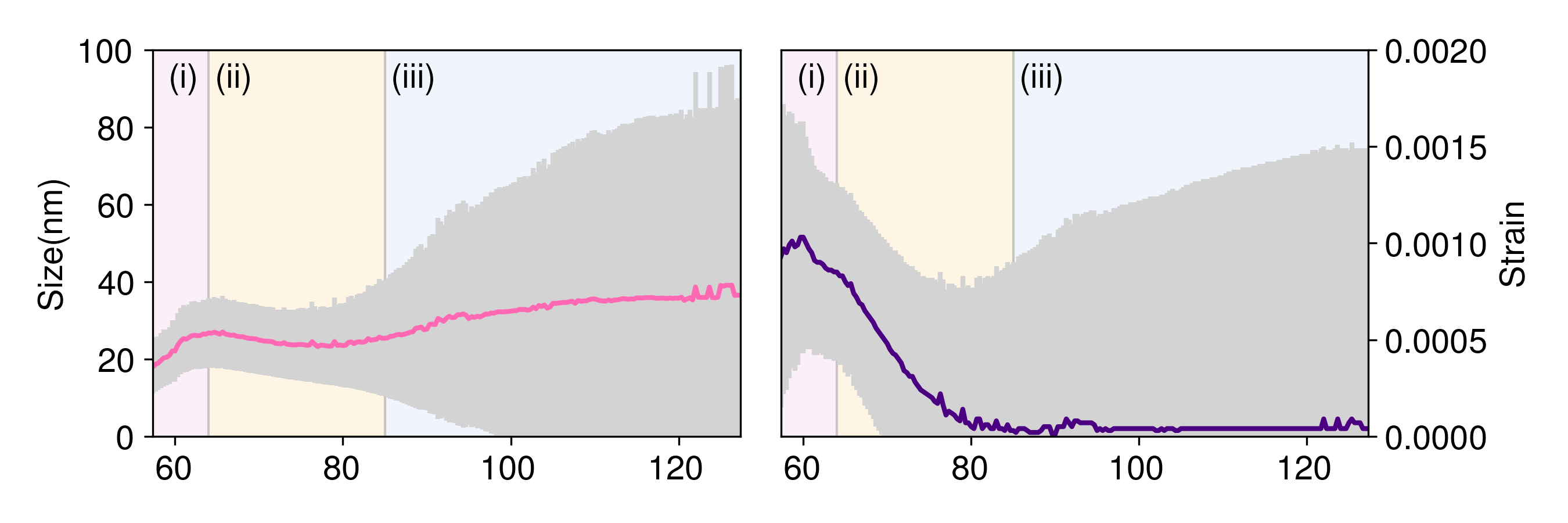}
\caption{\label{S-fig:particlesizeY2O3} The calculated size and strain values of the coherently diffracting domains of \ce{Y2O3} intermediates as calculated from the Scherrer equation with error bars using the Integral Breadth method for peak width. The colored regions (i,ii,iii) are taken from Figure~\ref{fig:insitucationplot} and Figure~\ref{fig:fstarP3}.}
\end{center} 
\end{figure}

Rietveld analysis of the synchrotron X-ray diffraction data presented in Figure~\ref{fig:insitucationplot} provides a quantitative relationship between all crystalline phases that occur at significant phase fractions ($>$ ca.~1\%).  
The first  step that is observed along the pathway is formation of yttrium oxychloride and sodium chloride from the reaction between \ce{YCl3} and \ce{Na2CO3} at 230 \textcelsius{}: 
\begin{align}
\label{S-eq:YOCl}
\cee{& YCl3 + Na2CO3 -> YOCl + 2NaCl + CO2},
\end{align}
\noindent The formation of NaCl at 230 \textcelsius{} corresponds directly with the increase in intensity of the $P4/mnm$ YOCl phase and loss of \ce{Na2CO3} phase intensity, consistent with lithium carbonate assisted metathesis reactions.\cite{Todd2019a}   YOCl  persists over the temperature range  260-490 \textcelsius{}, reacting with more sodium carbonate to form \ce{Y3O4Cl} at 450 \textcelsius{}, and eventually yielding \ce{Y2O3} at 500 \textcelsius{}.  Notably, the particle size of \ce{Y2O3} remains small (5-40 nm) until its subsequent consumption; the particle microstrain is insignificant in small in all cases (Figure~\ref{S-fig:particlesizeY2O3}). 

The precursor \ce{Mn2O3} remains unreacted until 540 \textcelsius{} when remaining \ce{Na2CO3} reacts to form the P3-\ce{Na_{$x$}MnO2} intermediate: 
\begin{align}
\label{S-eq:NaxMnO2}
\cee{$x$Na2CO3 + Mn2O3 ->  2 \text{P3-}Na_{$x$}MnO2 + $x$CO2},
\end{align}
consistent with other studies.\cite{Bianchini2020}
The P3-\ce{Na_{$x$}MnO2} intermediate is best described by the monoclinic \emph{C2/m} P3 layering observed in analogous sodium cobalt oxide structures. \cite{Takada2005,Bianchini2020} As the P3-\ce{Na_{$x$}MnO2} reacts with \ce{Y2O3} to form the pyrochlore, P2$^\prime$-\ce{Na_{$x$}MnO2} also forms. This P2$^\prime$-\ce{Na_{$x$}MnO2} intermediate forms in the $Cmcm$ structure where a P2$^\prime$ designation occurs from a slight shift of the  \ce{$M$O2} layers to cause a monoclinic distortion from the idealized P2-$P6_{3}/mmc$ structure. In P2$^\prime$-\ce{Na_{$x$}MnO2},  sodium occupies both the prismatic and octahedral sites at fractional occupancy.\cite{Caballero2002, Co1999, Clement2015}

\begin{figure}[h]
\begin{center}
\includegraphics[width=5in]{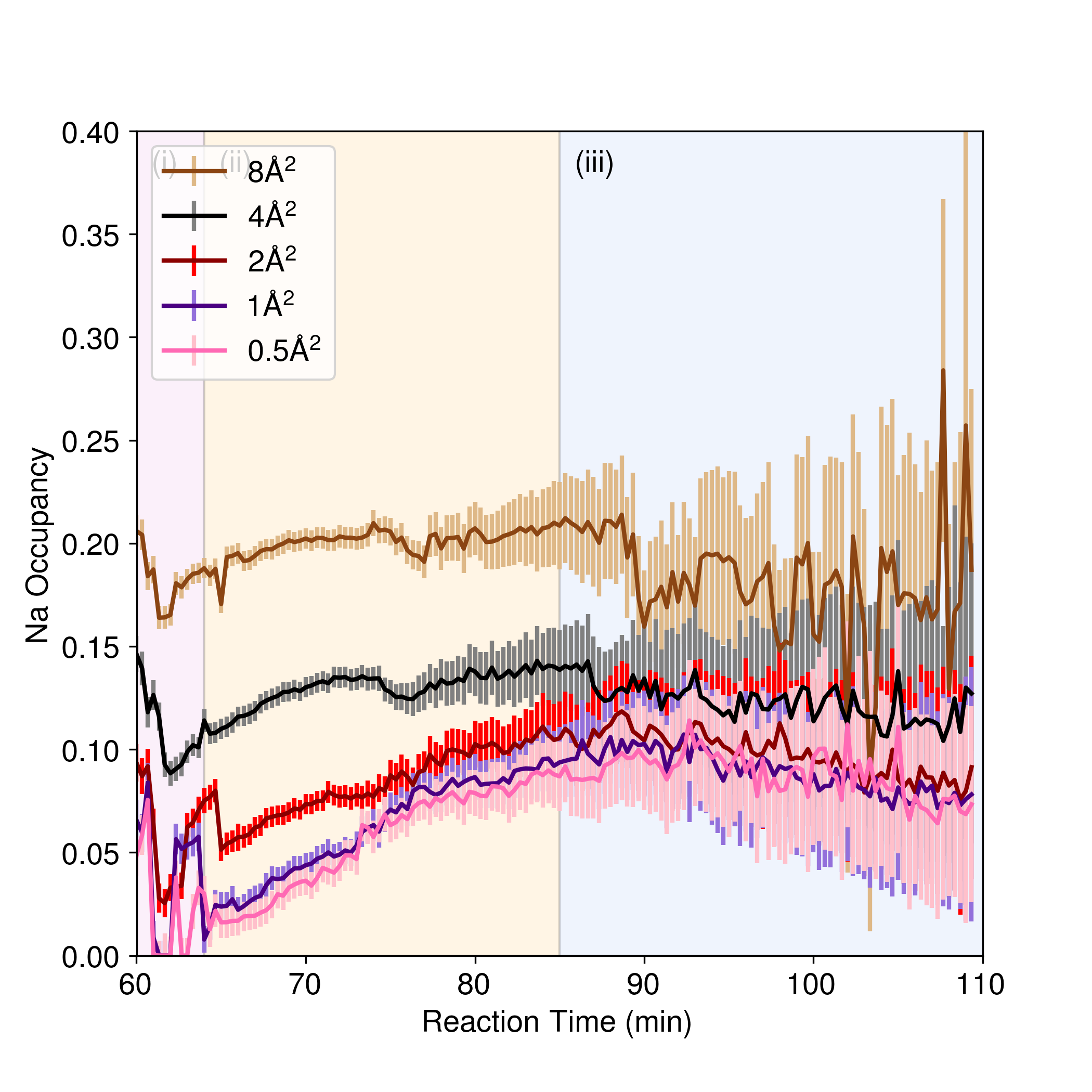}
\caption{\label{S-fig:P3NaOccvsBeq} Calculated sodium site occupancy values from sequential Rietveld refinements for the \emph{C2/m} \ce{Na_{x}MnO2} intermediate as a function of reaction time at different fixed thermal displacement parameters. The 2 \AA$^2$ value was used in calculating the $f^{*}$ ternary diagrams in Figure~\ref{fig:fstarP3}. The colored regions (i,ii,iii) are taken from Figure~\ref{fig:insitucationplot}.}
\end{center} 
\end{figure}

\begin{figure}[h]
\begin{center}
\includegraphics[width=5in]{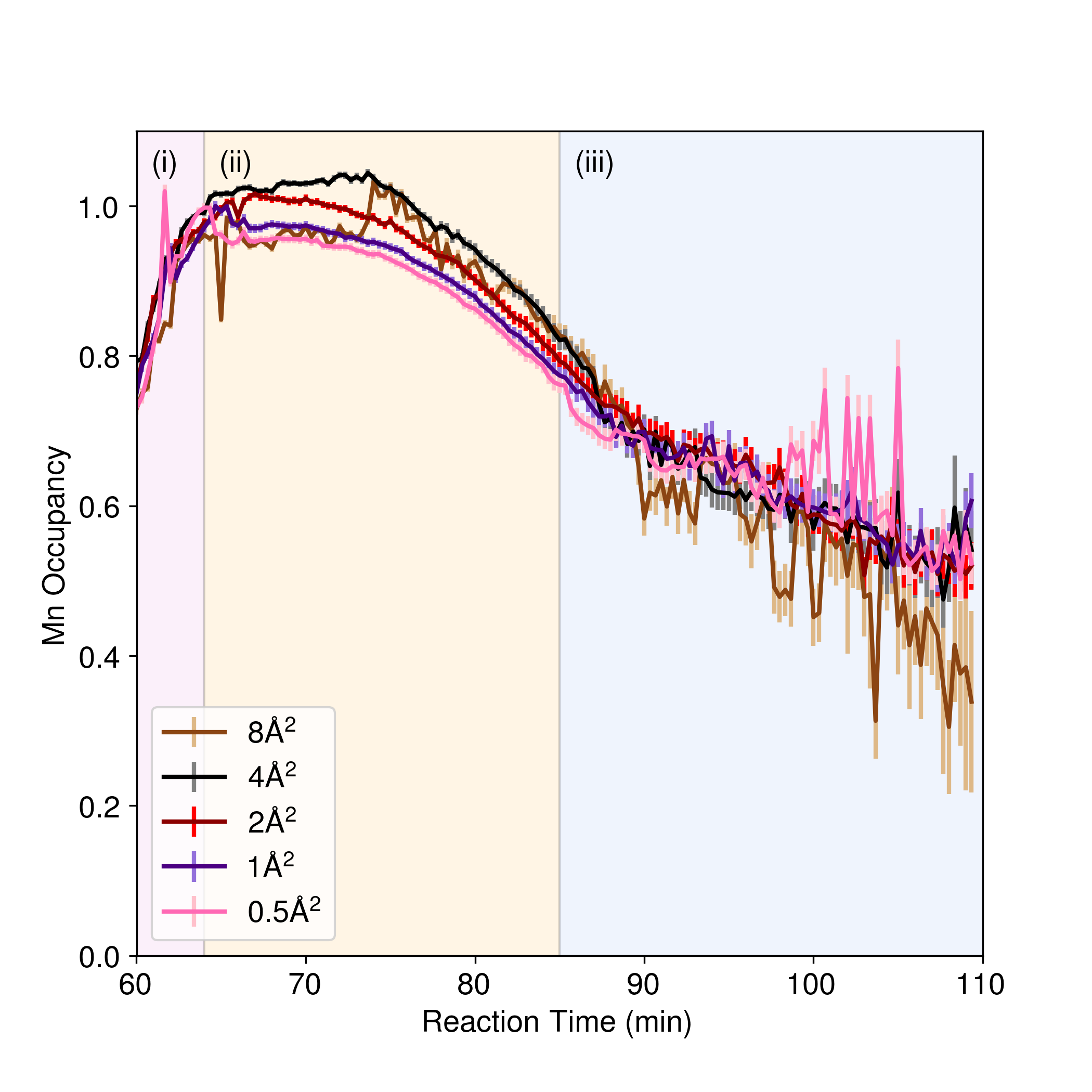}
\caption{\label{S-fig:P3MnOccvsBeq} Calculated manganese site occupancy values from sequential Rietveld refinements for the \emph{C2/m} \ce{Na_{x}MnO2} intermediate as a function of reaction time at different fixed thermal displacement parameters. The 2 \AA$^2$ value was used in calculating the $f^{*}$ ternary diagrams in Figure~\ref{fig:fstarP3}. The colored regions (i,ii,iii) are taken from Figure~\ref{fig:insitucationplot}.}
\end{center} 
\end{figure}

\begin{figure}[h]
\begin{center}
\includegraphics[width=5in]{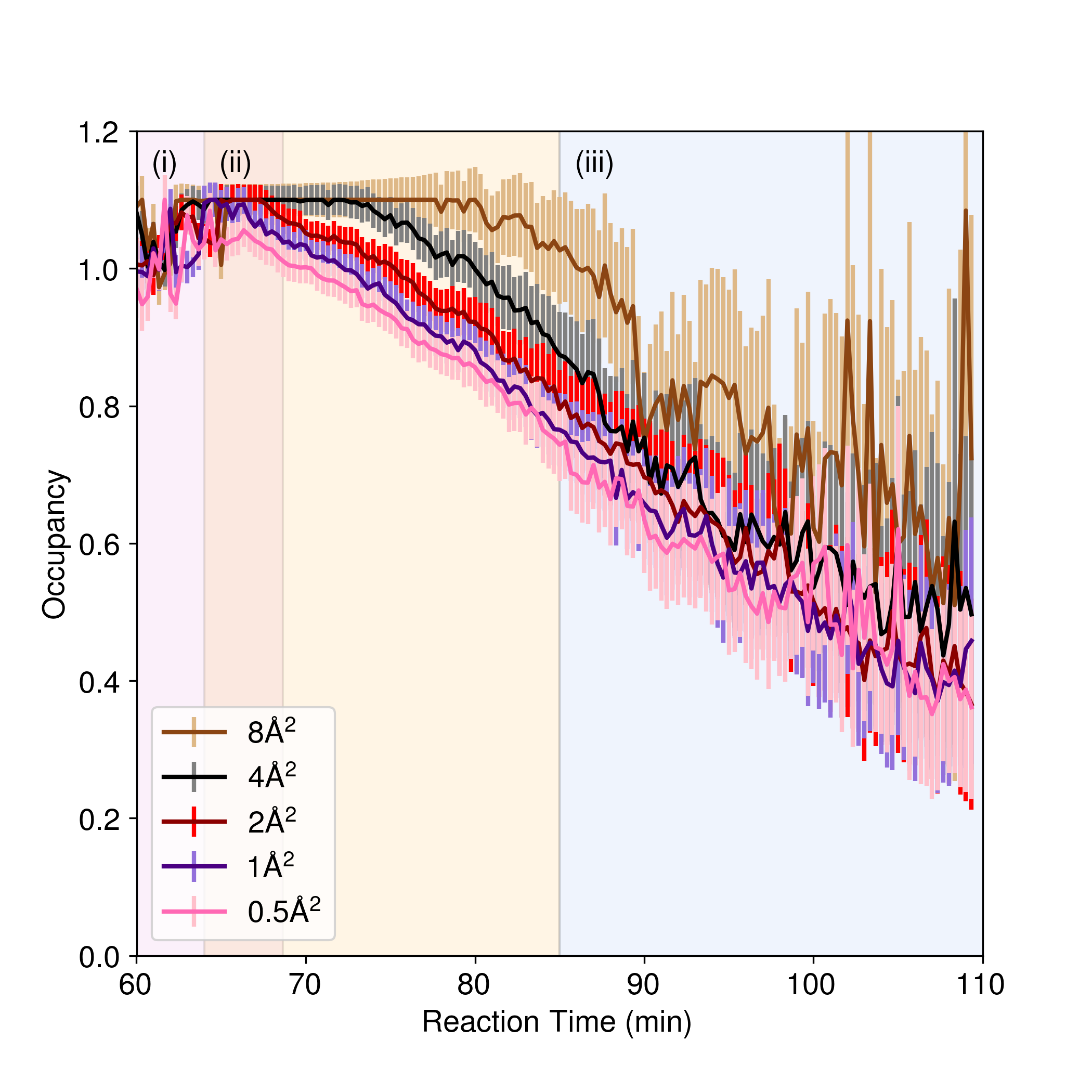}
\caption{\label{S-fig:P3OOccvsBeq} Calculated oxygen site occupancy values from sequential Rietveld refinements for the \emph{C2/m} \ce{Na_{x}MnO2} intermediate as a function of reaction time at different fixed thermal displacement parameters. The 2 \AA$^2$ value was used in calculating the $f^{*}$ ternary diagrams in Figure~\ref{fig:fstarP3}. The colored regions (i,ii,iii) are taken from Figure~\ref{fig:insitucationplot}.}
\end{center} 
\end{figure}

The initial formation of P3-\ce{Na_{$x$}MnO2} proceeds through a cation-rearrangement reaction where manganese occupying the ($8j$) sodium site moves to the manganese (2a) site. This is described by following the compositional evolution during the reaction of each crystallographic site on a $f^{*}$ diagram. \cite{Yin2018}  To avoid any correlations between atomic displacement parameters and crystallographic site occupancies, we performed many different sequential series of Rietveld refinements with different values for the $B_{iso}$ atomic displacement parameters (Figures~\ref{S-fig:P3NaOccvsBeq}-\ref{S-fig:P3OOccvsBeq}); the ADP's were held constant to avoid covariances, especially with the multiphased refinement.  Using a $B_{iso}$ = 2~\AA$^2$ yielded the most reproducible refinements, lowest statistical errors, and small covariances of refined parameters.  

To interpret the $f^{*}$ diagram, trajectory (i) in Figure~\ref{fig:fstarP3} describes the defect rearrangement reaction: \begin{align}
\label{S-eq:cationswap}
\cee{\text{P3-}[Na_{1-y}Mn_{z}][Mn_{1-x}]O2 -> \text{P3-}[Na_{1-y}][Mn_{1-x+z}]O2},
\end{align}
\noindent where the first set of brackets describes the composition of the ideal sodium sites, and the second set of brackets describes that of manganese. The position of the starting composition is calculated as P3-\ce{Na_{0.35}Mn_{0.79}O2}.  At its  maximal phase fraction  observed in Figure~\ref{fig:insitucationplot}, the composition is closer to \ce{Na_{0.28}Mn_{0.96}O_{2}}.  At that point in the overall reaction, this intermediate reacts to form \ce{Y2Mn2O7}.

\begin{figure}[h]
\begin{center}
\includegraphics[width=5in]{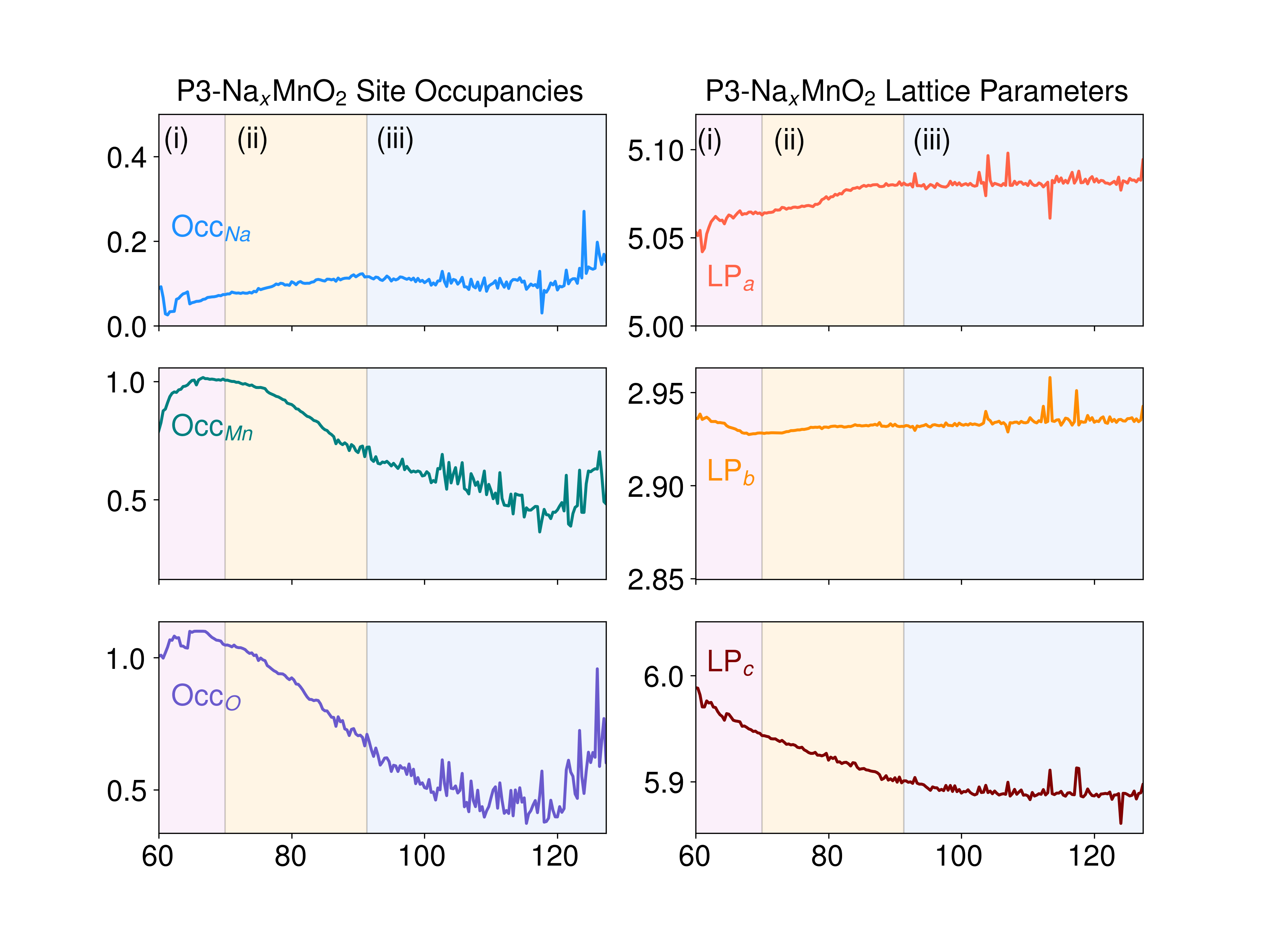}
\caption{\label{S-fig:P3parameters} Structural parameters for the \emph{C2/m} \ce{Na_{x}MnO2} intermediate as a function of reaction time. Refined occupancies of each atomic site that were used in calculating the $f^{*}$ ternary diagrams are shown on the right while Unit cell parameters for the monoclinic cell are shown on the left. The colored regions (i,ii,iii) are taken from Figure~\ref{fig:insitucationplot} and Figure~\ref{fig:fstarP3}.}
\end{center} 
\end{figure}

As \ce{Y2Mn2O7} starts to form (Figure~\ref{fig:insitucationplot}), there is a change in the defect chemistry of P3-\ce{Na_{$x$}MnO2}.  This is indicated by trajectory (ii) in Figure~\ref{fig:fstarP3} that ends at \ce{Na_{0.65}MnO2}. This result is observed from the decrease in relative electron density from the oxygen site (4i), as well as an increase in the sodium to manganese site ratio. This trend suggests that sodium is either intercalated back into the structure, or that the intermediate loses manganese and oxygen. Figure~\ref{S-fig:P3parameters} supports the latter, where the occupancy of both manganese and oxygen decreases as \ce{Y2Mn2O7} forms, while the interlayer spacing between manangese oxide layers decreases. This provides the balanced reaction: 
\begin{align}
\label{S-eq:simulreactions}
\cee{4\text{P3-}Na_{0.25}MnO2 + Y2O3 -> Y2Mn2O7 + 2\text{P2$^\prime$-}Na_{0.5}MnO2}.
\end{align}
During the process described in Eqn.~\ref{S-eq:simulreactions}, the pyrochlore grows rapidly (Figure~\ref{fig:insitucationplot}, 70--90 min) before tapering off at the end of the second trajectory in Figure~\ref{fig:fstarP3} as the amount of P3-\ce{Na_{$x$}MnO2} depletes. 

\begin{figure}[h]
\begin{center}
\includegraphics[width=5in]{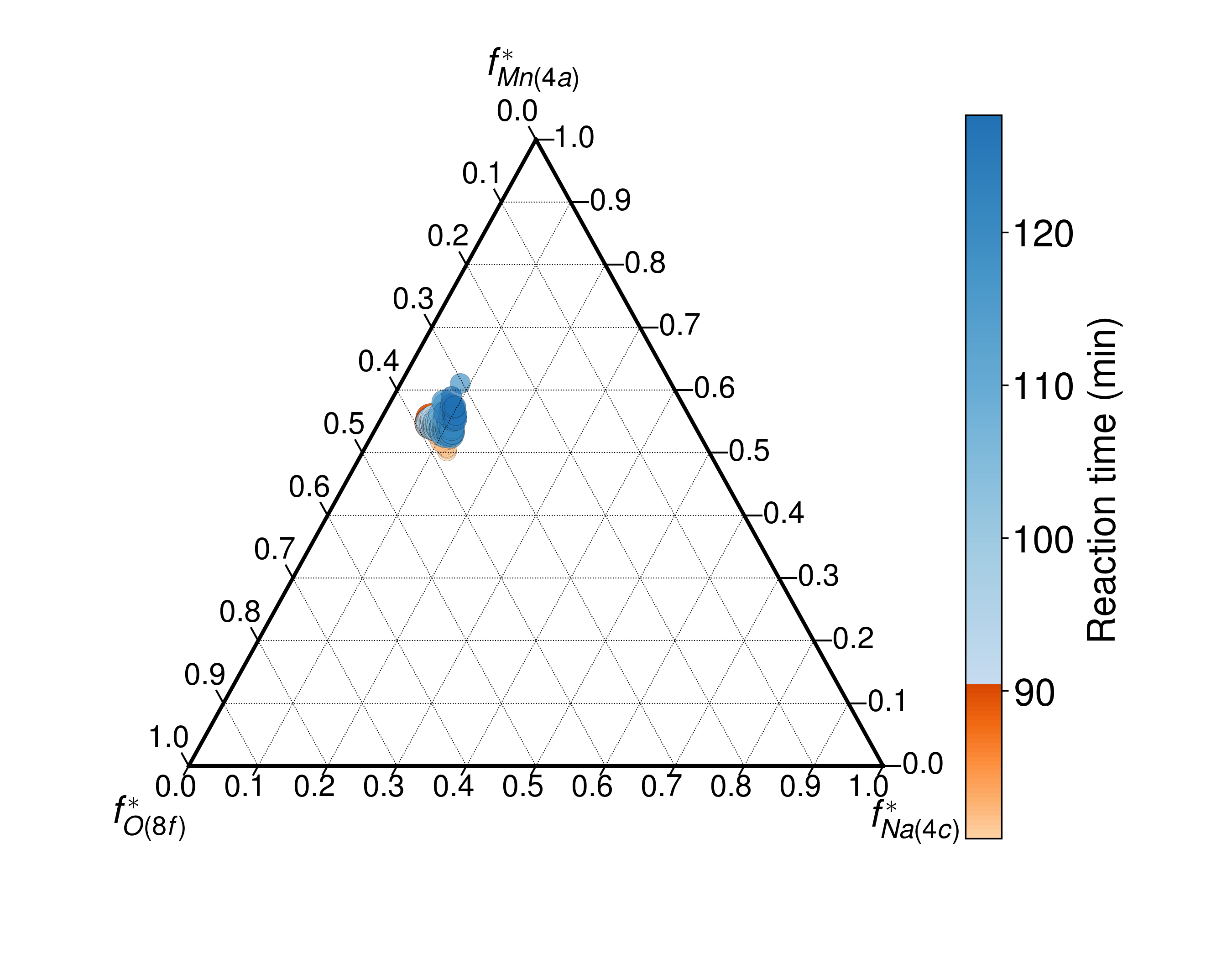}
\caption{\label{S-fig:fstarP2} Ternary $f^{*}$ diagram of atomic coordinates calculated from Rietveld refinement of $Cmcm$ \ce{Na_{x}MnO2}. Each axis denotes atom identity and crystallographic Wyckoff position within the lattice. The color bar in the figure is separated into the two defect reactions where the P2$^\prime$ phase is present, as depicted in Figure~\ref{fig:insitucationplot}. The diameter of each circle mirrors the calculated Weighted Scale Factor in Figure~\ref{fig:insitucationplot} for P2$^\prime$-\ce{Na_{x}MnO2}.}
\end{center} 
\end{figure}

\begin{figure}[h]
\begin{center}
\includegraphics[width=5in]{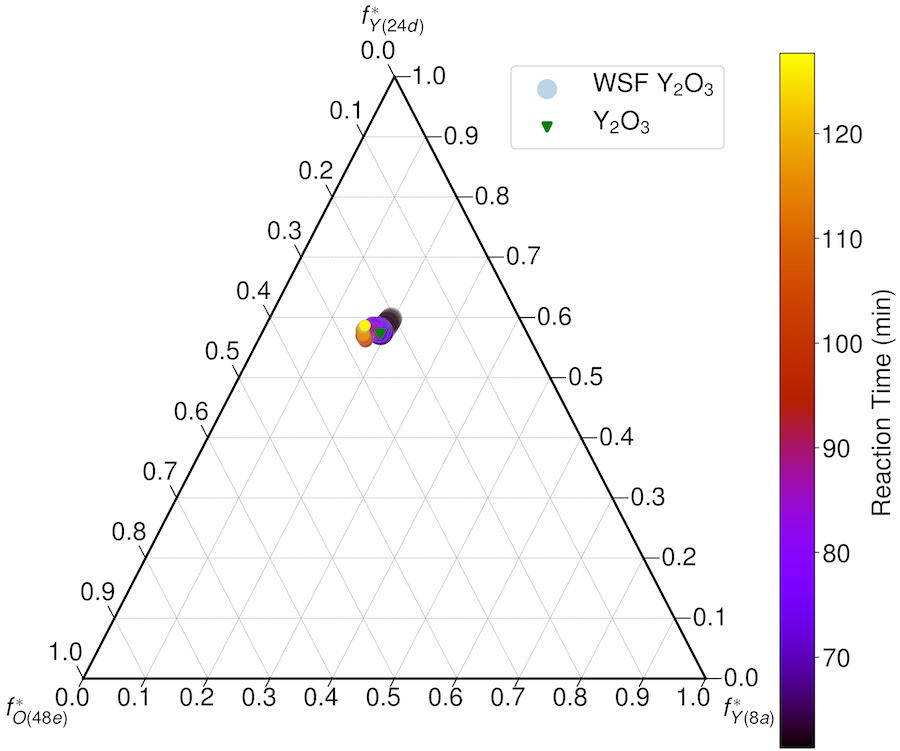}
\caption{\label{S-fig:fstarY2O3} Ternary $f^{*}$ diagram of atomic coordinates calculated from Rietveld refinement of $Ia\overline{3}$ \ce{Y2O3}. Each axis denotes atom identity and crystallographic Wyckoff position within the lattice. For reference, the calculated value of \ce{Y2O3}  is provided. The color bar in the figure follows the lifetime of the \ce{Y2O3} intermediate in the assisted metathesis reaction presented in Figure~\ref{fig:insitucationplot}. The diameter of each circle mirrors the calculated Weighted Scale Factor in Figure~\ref{fig:insitucationplot} for \ce{Y2O3}.}
\end{center} 
\end{figure}

Formation of \ce{Y2Mn2O7} is sustained through the reaction of P2$^\prime$-\ce{Na_{x}MnO2}. Compositional analysis of the P2$^\prime$-\ce{Na_{$x$}MnO2} phase on an $f^{*}$ diagram (Figure~\ref{S-fig:fstarP2}) reveals that the stoichiometry does not deviate systematically away from \ce{Na_{0.5}MnO2} during the continued course of reaction.  Furthermore, the composition of \ce{Y2O3} does not change during the course of the reaction (Figure~\ref{S-fig:fstarY2O3}).  The consumption of P2$^\prime$-\ce{Na_{0.5}MnO2} also includes the consumption of \ce{Y2O3} with production of both \ce{Y2Mn2O7} and \ce{Na3MnO4} (Figure~\ref{fig:insitucationplot}), as described in the balanced reaction:
\begin{align}
\label{S-eq:NaMn2O4}
\cee{12\text{P2$^\prime$-}Na_{0.5}MnO2 + 5 Y2O3 + 8 O2 -> 5 Y2Mn2O7 + 3 Na3MnO4}.
\end{align}
This helps to explain the need for excess (flowing) oxygen in the previous study of this reaction.\cite{Todd2019}.

\section{Testing of alternative NaMnO$_2$ polytypes}
To determine if the stacking ordering of the manganese-oxygen layers in \ce{NaMnO2} polymorphs affects the observed products, $C2/m$ \ce{NaMnO2} was prepared. This monoclinic polymorph adopts an O3 stacking variant as compared to the P2$^\prime$ $Cmcm$ polymorph. The $C2/m$ polytype differs from the P2$^\prime$ variant by a shift of the layers, and a 60$^{o}$ rotation of all [\ce{MnO6}] octahedra. PXRD results starting with the monoclinic O3 polymorph are presented in Figure~\ref{S-fig:mNaMnO2YOCl}.

\begin{figure}[h]
\begin{center}
\includegraphics[width=6.in]{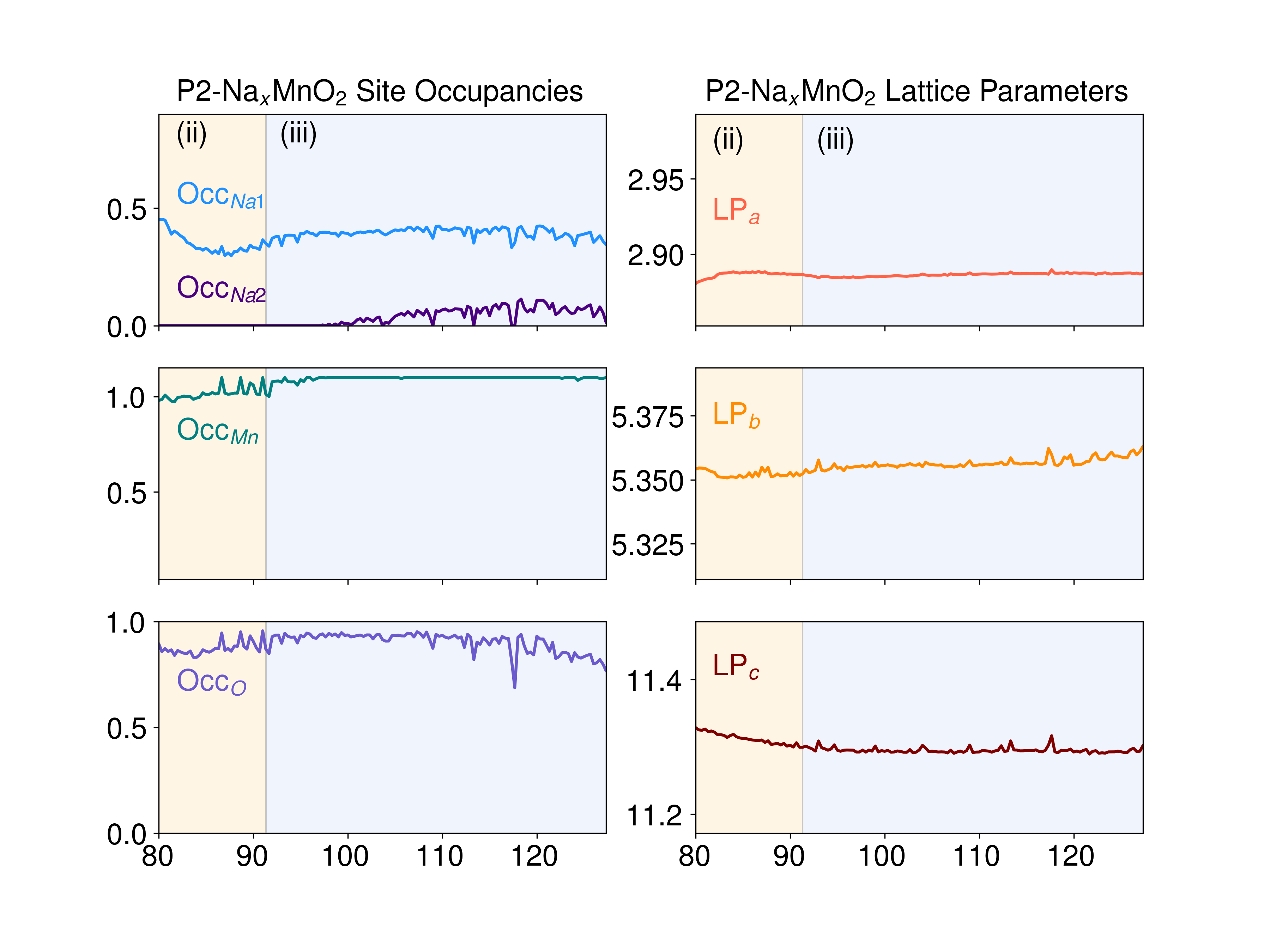}
\caption{\label{S-fig:P2parameters} Structural parameters for the $Cmcm$ \ce{Na_{x}MnO2} intermediate as a function of reaction time. Refined occupancies of each atomic site that were used in calculating the $f^{*}$ ternary diagrams are shown on the right while Unit cell parameters for the orthorhombic cell are shown on the left. The colored regions (i,ii,iii) are taken from Figure~\ref{fig:insitucationplot} and Figure~\ref{S-fig:fstarP2}.}
\end{center} 
\end{figure}

The calculated occupancy of sodium sites in the lattice supports the proposed mechanism to remove [\ce{MnO2}] species. The $Cmcm$ structure has two unique interstitial sodium sites that relate to their position relative to [\ce{MnO6}] octahedra. The Na1 site is denoted as the octahedral site, as it sits above and below the [\ce{MnO6}] octahedral faces, while the Na2 site shares edges with the [\ce{MnO6}] octahedra. The distance between these sodium sites is $\sim1.6$~\AA, thus, both sites cannot be occupied simultaneously, and a fractional occupancy is observed at higher sodium occupancies. The sodium occupancy used in the $f^{*}$ ternary maps is the summed value of both site occupancies, whereas the individual values are plotted in Figure~\ref{S-fig:P2parameters}. Interestingly, upon formation of \ce{Na_{$x$}MnO2} during process (i) (Figure~\ref{fig:fstarP3}, Eqn.~\ref{S-eq:cationswap}), only the octahedral Na1 site is occupied, an observation that is supported in P2-type \ce{Na_{x}MnO2} materials with a high degree of cation defects in the structure (e.g., cathode material NMO-420).\cite{Stoyanova2010}. As these defects are removed during the reaction pathway, the Na2 site populates to reduce electrostatic repulsion. The $f^{*}$ map suggests that during trajectory (ii), eventually, the occupancy of sodium approaches \ce{NaMnO2}, which is unusual for P2-type materials where sodium occupancy hits a maximum at 0.7 before a phase transition is observed to O3 type ordering as seen electrochemically.\cite{Clement2015} This transition is not observed in the diffraction data here suggesting that a higher amount of sodium can be stabilized in the P2 structure at higher temperatures or under non-equilibrium conditions.

\section{\textit{Ex situ} control experiments}

\begin{figure}[h]
\begin{center}
\includegraphics[width=5in]{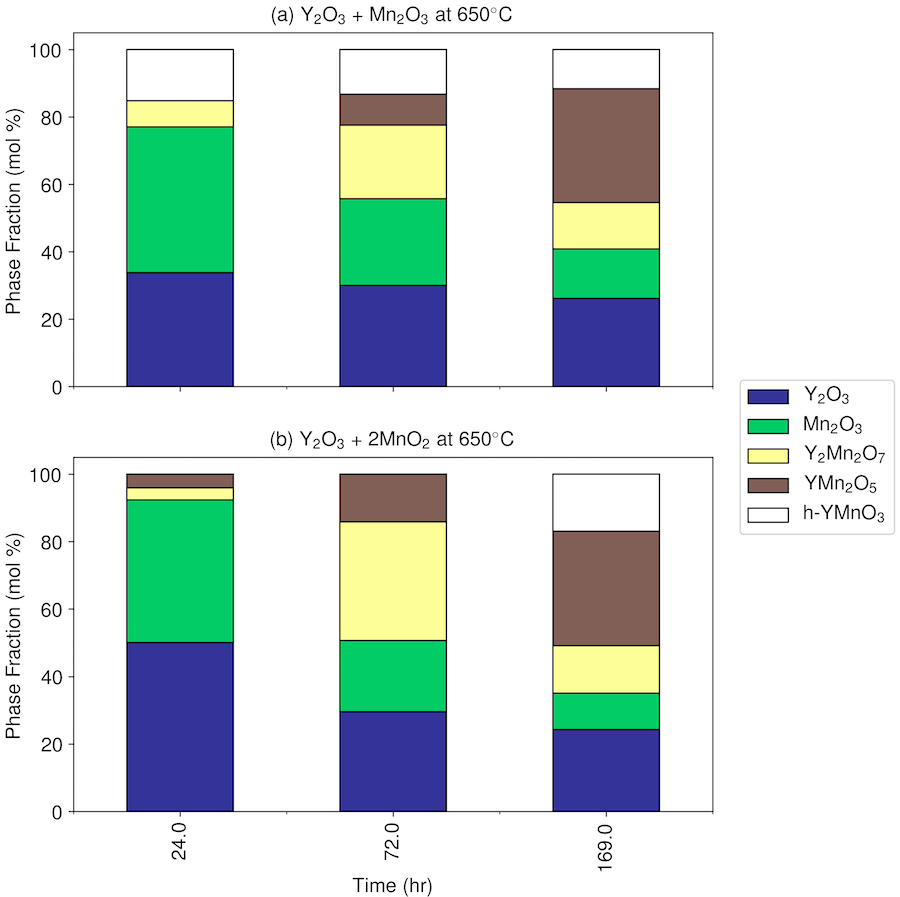}
\caption{\label{S-fig:MnOControls} Bar graphs representing the mole percent amounts of products calculated from $ex$ $situ$ PXRD Rietveld refinements of the reaction: (a) \ce{Mn2O3 + Y2O3 ->[O_{2}]} and (b) \ce{2MnO2 + Y2O3 ->[O_{2}]}. The reactions were heated at 10 \textcelsius{}/min to 650 \textcelsius{} for 24, 72, and 128 hours under flowing oxygen. Each color bar is represented in the legend as a different product phase.}
\end{center} 
\end{figure}

Control reactions of binary yttrium and manganese oxides reveal that the sodium carbonate assisted metathesis reaction pathway is selective for \ce{Y2Mn2O7}. Analysis of \textit{ex situ} control reactions, \ce{2MnO2 + Y2O3} and \ce{Mn2O3 + Y2O3}, performed at 650 \textcelsius{} under flowing \ce{O2} at ambient pressure (Figure~\ref{S-fig:MnOControls}) reveal that the reactions contain a finite fraction of \ce{Y2Mn2O7} at initial time points (i.e., no selectivity), and the phase fraction of \ce{Y2Mn2O7} decreases at longer times in favor of \ce{YMn2O5}. Based on Figure~\ref{fig:YMnOPhaseDiagram} this suggests that the true $\mu_{O}$ is near the border between \ce{YMn2O5} and \ce{Y2Mn2O7}. 

\begin{figure}[h]
\begin{center}
\includegraphics[width=5in]{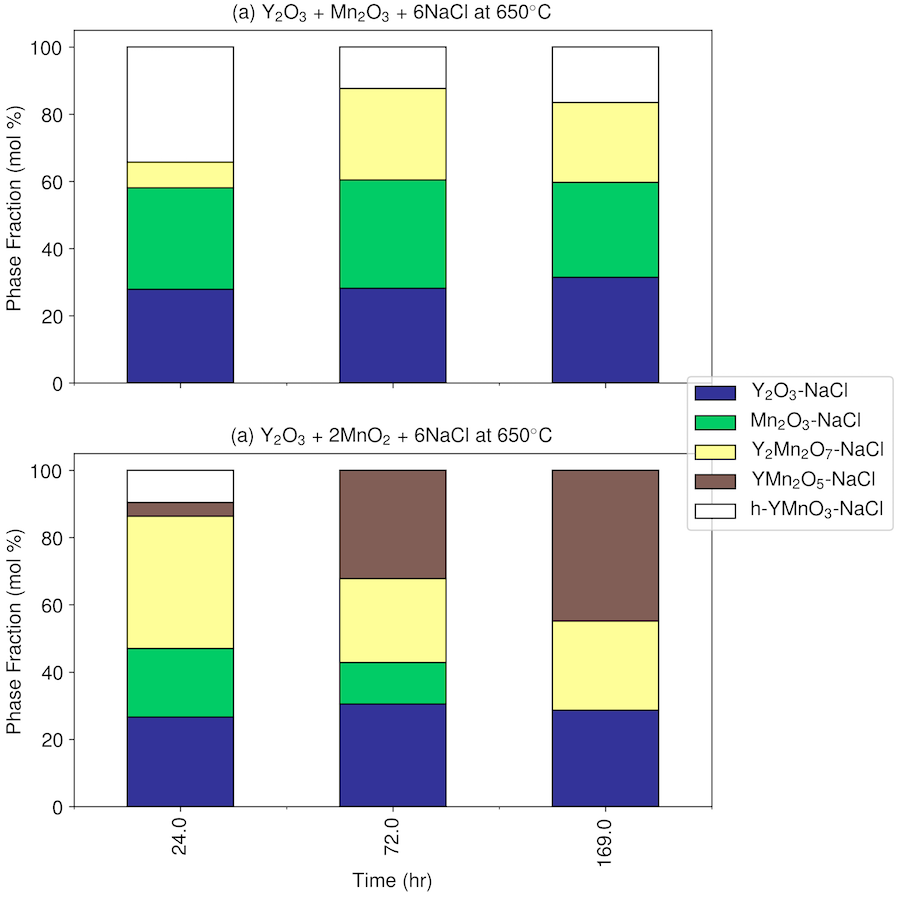}
\caption{\label{S-fig:MnONaClControls} Bar graphs representing the mole percent amounts of products calculated from $ex$ $situ$ PXRD Rietveld refinements of the reaction: (a) \ce{Mn2O3 + Y2O3 ->[6NaCl, O_{2}]} and (b) \ce{2MnO2 + Y2O3 ->[6NaCl, O_{2}]}. The reactions were heated at 10 \textcelsius{}/min to 650 \textcelsius{} for 24, 72, and 128 hours under flowing oxygen. Each color bar is represented in the legend as a different product phase. Mole fractions are presented after subtracting the contribution from NaCl for comparison.}
\end{center} 
\end{figure}

\begin{figure}[h]
\begin{center}
\includegraphics[width=5in]{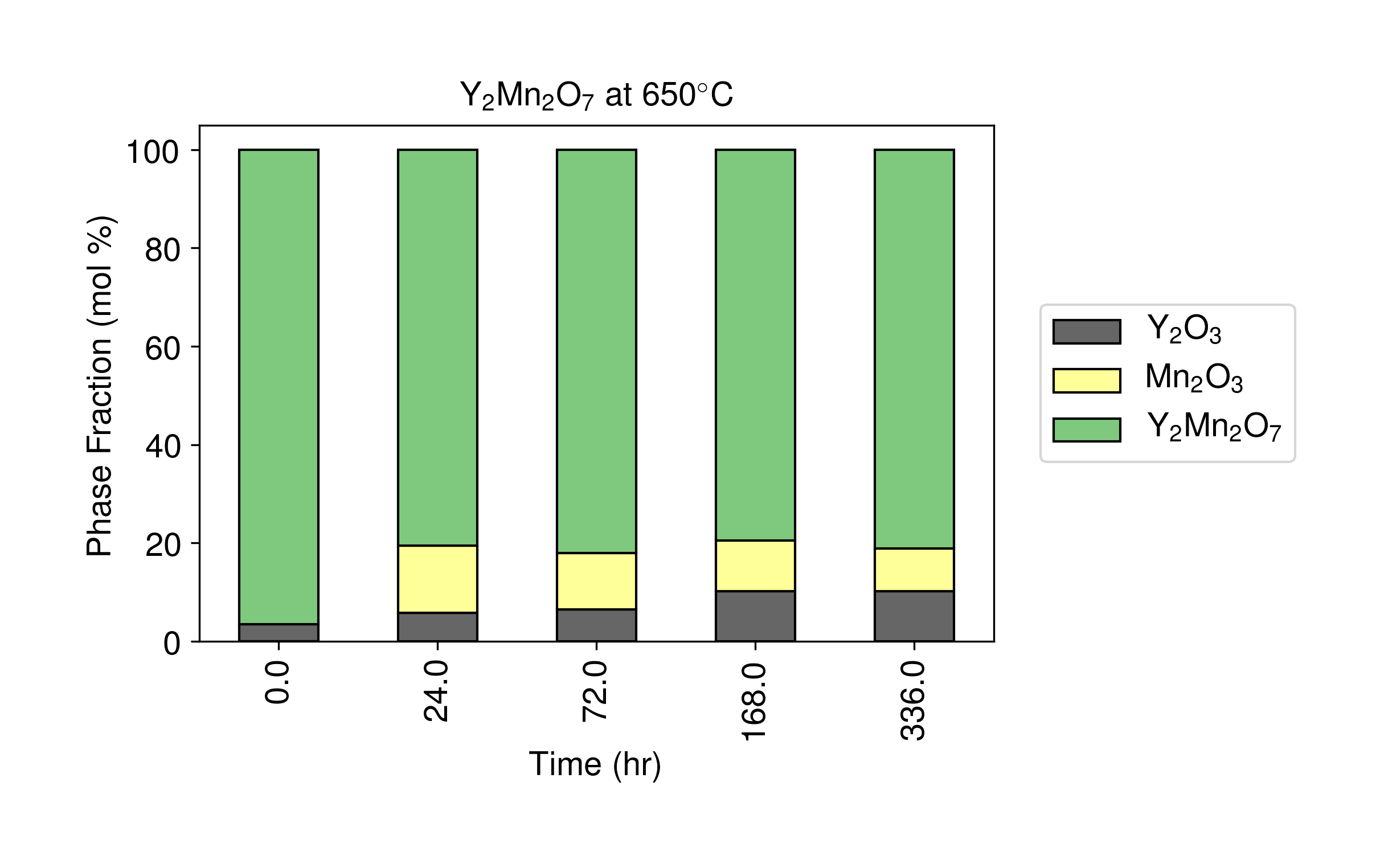}
\caption{\label{S-fig:Y2Mn2O7} Bar graphs representing the mole percent amounts of products calculated from $ex$ $situ$ PXRD Rietveld refinements of \ce{Y2Mn2O7}. The reactions were heated at 10 \textcelsius{}/min to 650 \textcelsius{} for 24, 72, 128, and 336 hours under flowing oxygen. Each color bar is represented in the legend as a different calculated phase.}
\end{center} 
\end{figure}

Additionally, the presence of a six molar equivalent of sodium chloride does increase the initial yield of \ce{Y2Mn2O7}, but \ce{YMn2O5} increases in phase fraction at long reaction times  (Figure~\ref{S-fig:MnONaClControls}). In order to assess the stability of pyrochlore, washed \ce{Y2Mn2O7} was heated at 650 \textcelsius{} under flowing oxygen for long-duration reactions.  Only partial decomposition occurs to the binary oxides, \ce{Mn2O3 + Y2O3}, after 24 h; the composition remains constant even after heating for 2 weeks (Figure~\ref{S-fig:Y2Mn2O7}). 

\begin{figure}[h]
\begin{center}
\includegraphics[width=5in]{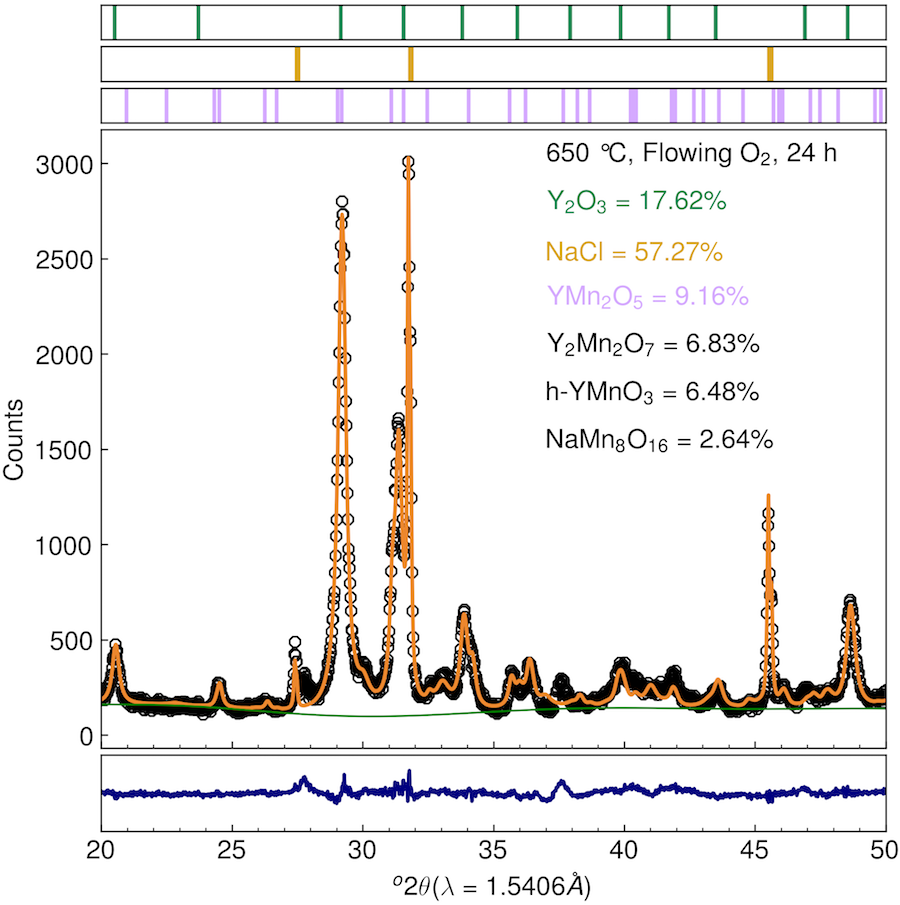}
\caption{\label{S-fig:mNaMnO2YOCl} PXRD results from the reaction: $C2/m$ \ce{NaMnO2 + YOCl ->[O_{2}]}. The reaction was performed by heating at 10 \textcelsius{}/min to 650 \textcelsius{} and dwelling for 24 h. Raw data: black circles, Rietveld refinement: orange, background: green, difference: navy. Tick marks denote the majority phases and are highlighted by color and mole percent as calculated from Rietveld refinements. All other product mole percents are highlighted in black without tick marks.}
\end{center} 
\end{figure}

\begin{figure}[h]
\begin{center}
\includegraphics[width=5in]{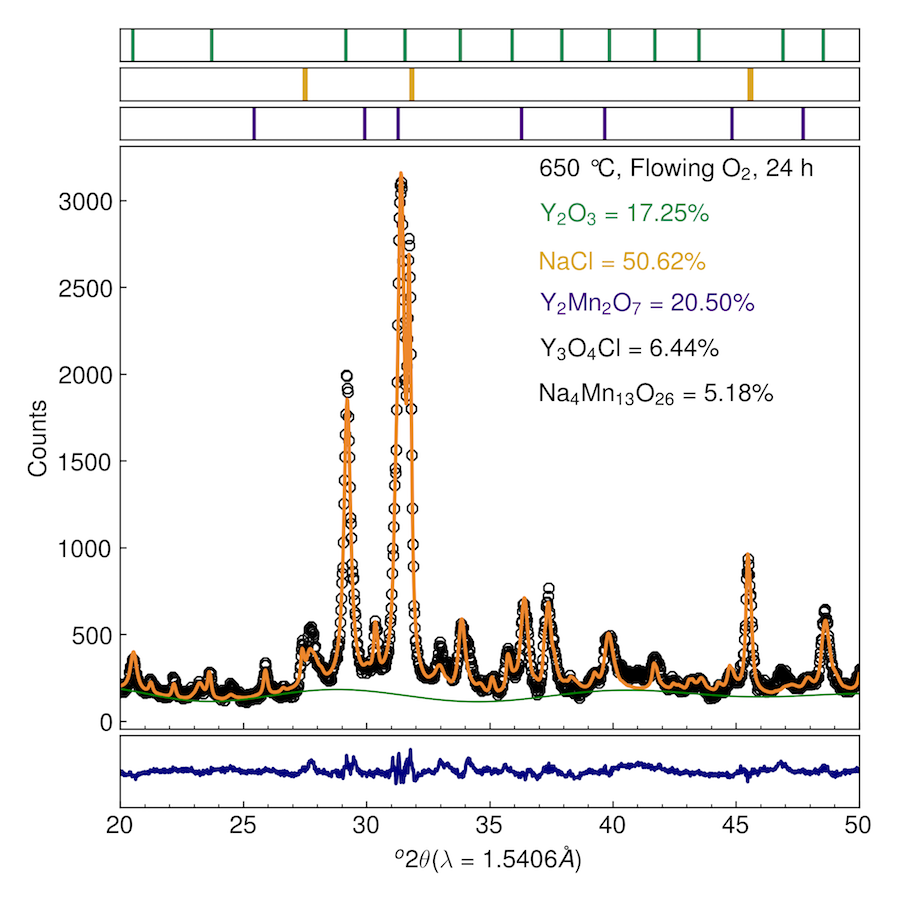}
\caption{\label{S-fig:oNaMnO2YOCl} PXRD results from the reaction: $Cmcm$ \ce{Na_{0.7}MnO2 + YOCl ->[O_{2}]}. The reaction was performed by heating at 10 \textcelsius{}/min to 650 \textcelsius{} and dwelling for 24 h. Raw data: black circles, Rietveld refinement: orange, background: green, difference: navy. Tick marks denote the majority phases and are highlighted by color and mole percent as calculated from Rietveld refinements. All other product mole percents are highlighted in black without tick marks.}
\end{center} 
\end{figure}

Control ternary metathesis reactions performed using nominally stoichiometric, ideal sodium manganese oxide precursors do not yield selectivity for \ce{Y2Mn2O7}. In the case of \ce{Li2CO3}-based assisted metathesis reactions,\cite{Todd2019} we previously identified that the ternary metathesis reaction, \ce{LiMnO2 + YOCl -> YMnO3 + LiCl} yielded improved selectivity for orthorhombic \ce{YMnO3} over the analogous assisted metathesis reactions.\cite{Todd2019a}  However, the reaction, \ce{2NaMnO2 + 2YOCl ->[O$_{2}$]}, results in a mixture of Y-Mn-O products (Figure~\ref{S-fig:mNaMnO2YOCl}). Furthermore, using a partially-oxidized precursor in the reaction, P2-\ce{2Na_{0.7}MnO2} with 2 YOCl selectively yields \ce{Y2Mn2O7}, but it is incomplete after 24 h (Figure~\ref{S-fig:oNaMnO2YOCl}). Together, these control reactions lend support to the sodium-based reaction pathway in selectively yielding \ce{Y2Mn2O7}. 

\section{Comparing neighboring ternary oxide structures in chemical potential space}

\begin{figure}[ht]
\begin{center}
\includegraphics[width=6.5in]{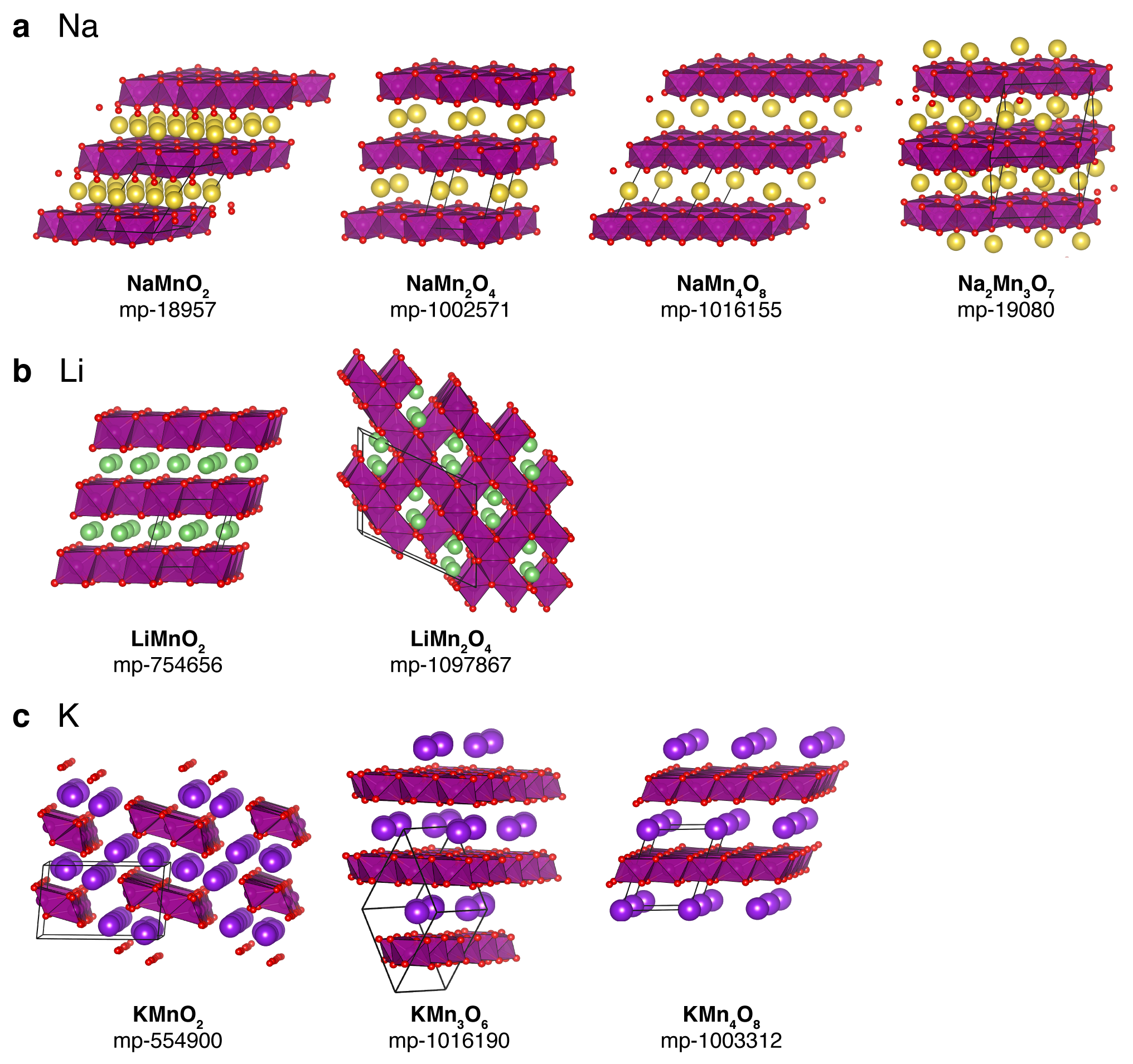}
\caption{\label{S-fig:structures} Structures of possible stoichiometric \textit{A}-Mn-O ternary oxide intermediates predicted to be stable at $T=650$ \textcelsius{} for \textit{A} equal to \textbf{a}, Na, \textbf{b}, Li, \textbf{c}, K. Unit cells are drawn and each structure's Materials Project ID (mp-id) is provided.}

\end{center} 
\end{figure}

Comparing the accessibility and similarity of ternary oxide structures offers insight into why the assisted metathesis reaction proceeds differently across the different alkali species. The structures of the possible stoichiometric \textit{A}-Mn-O ternary oxide intermediates are illustrated in Figure~\ref{S-fig:structures}. The sodium-based phases (Figure~\ref{S-fig:structures}a) show a strong preference for the layered Birnessite, $\delta$-\ce{MnO2}, framework. As illustrated by the gray shaded areas in Figure~\ref{fig:wormholes}b, the compositions of neighbors to \ce{NaMnO2} (mp-18957) that move towards higher $\mu_\text{O}$ include \ce{NaMn2O4} (mp-1002571), \ce{NaMn4O8} (mp-1016155) and \ce{Na2Mn3O7} (mp-19080). Note that the \ce{Na2Mn3O7} phase has a layered crystal structure akin to \ce{NaMnO2}, but with regularly ordered Mn vacancies; this phase is experimentally observed in trace quantities in 24 h reactions at 650 \textcelsius{} \cite{Todd2019}. For values of $x \leq 0.25$ in \ce{Na_{$x$}MnO2} (e.g., \ce{NaMn4O8}), the Hollandite $\alpha$-\ce{MnO2} framework was previously shown to be thermodynamically favorable \cite{Kitchaev2017}; however, the layered \ce{NaMn4O8} (mp-1016155) phase is predicted to be stable at $T=650$ \textcelsius{}.

In contrast, in the Li-Mn-O system (Figure~\ref{S-fig:structures}b) , as one looks at phases towards higher $\mu_{\text{O}}$ starting from the layered \ce{LiMnO2} (mp-754656), all neighboring phases would require a structural rearrangement (and thus kinetic barrier) as they become deficient in lithium. Nearly all of the subsequent phases prefer a spinel-like connectivity, which consists of three-dimensionally interconnected layers of edge-connected manganese-oxygen octahedra, including \ce{LiMn2O4} (mp-1097867, pictured), \ce{Li4Mn5O12} (mp-691115), and \ce{Li11Mn13O32} (mp-531021). A closer look at \ce{LiMn2O4} reveals that a polymorph with a layered structure is predicted (mp-1176656); however, it is 0.046 eV/atom above the hull (at $T=650$ \textcelsius{}).  

Finally, in the K-Mn-O system (Figure~\ref{S-fig:structures}c), the predicted ground state \ce{KMnO2} (mp-554900) phase is not of a layered \ce{MnO2} framework; although, the layered phase (mp-1002570) is about 0.044 eV/atom above the hull at $T=650$ \textcelsius{}. In the K-deficient regime, the layered phase dominates, but \ce{KMn2O4} (mp-1003314) is predicted to be metastable (0.007 eV/atom above hull). The layered structures \ce{KMn3O6} (mp-1016190) and \ce{KMn4O8} (mp-1003312) are both predicted to be stable, however. As in the Li-Mn-O system, accessing these K-deficient phases from the ground-state stoichiometric \ce{KMnO2} phase would require structural rearrangement, posing a significant kinetic barrier.  Experimentally, \ce{K2CO3}-based assisted metathesis reactions do yield some \ce{Y2Mn2O7} product at 650~\textcelsius{}; however, the reaction is not selective for \ce{Y2Mn2O7} \cite{Todd2019}, which is consistent with this analysis.

\section{Pareto frontier of predicted reactions forming \ce{Y2Mn2O7}}

\begin{figure}[ht!]
\begin{center}
\includegraphics[width=5in]{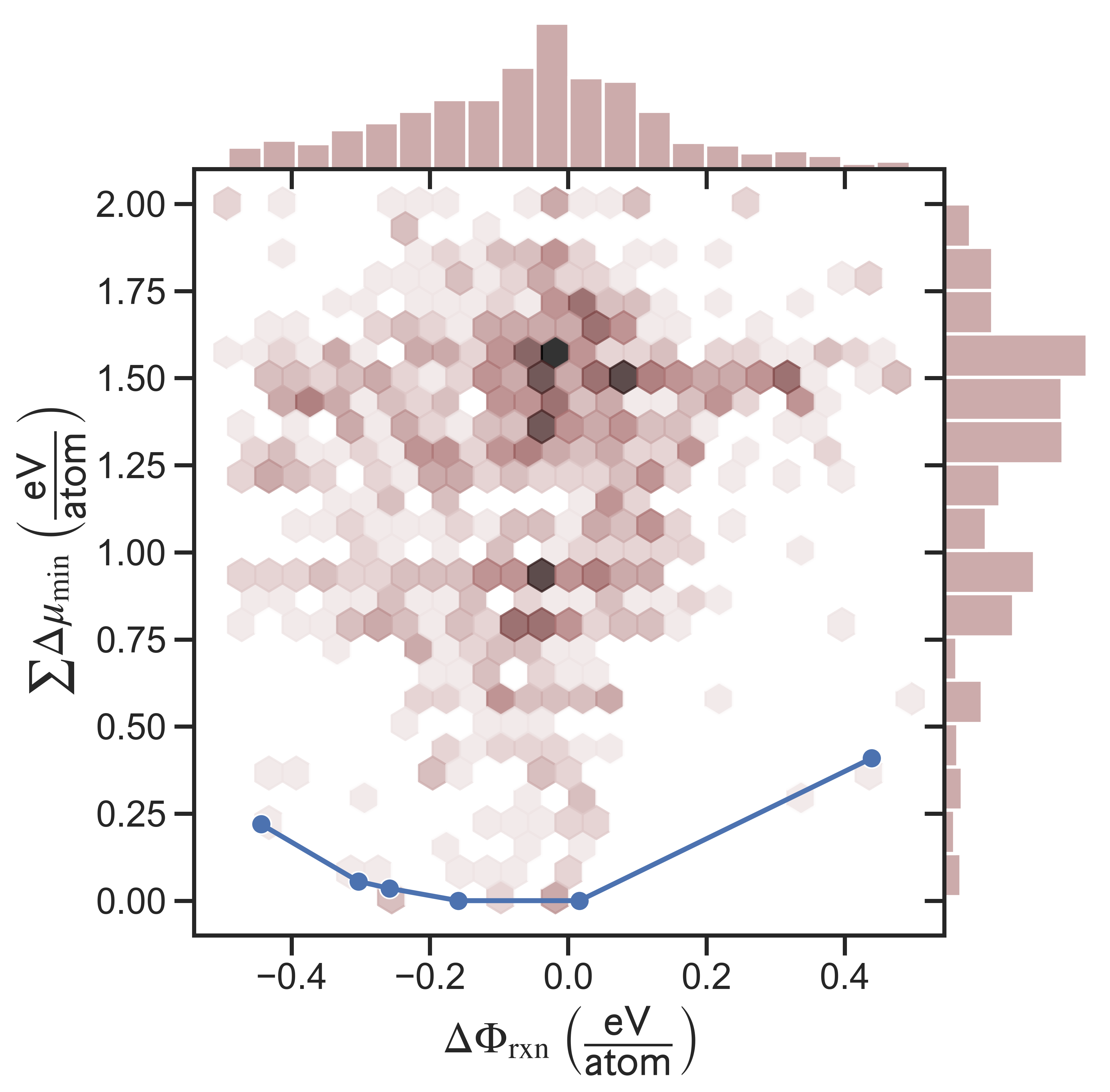}
\caption{\label{S-fig:pareto} \textbf{Energy and total chemical potential distance distribution for 3,017 predicted chemical reactions forming \ce{Y2Mn2O7}}. The Pareto frontier is given by the blue line. Reactions along this frontier, shown as blue dots, are reproduced in Table \ref{S-tab:pareto}.}

\end{center} 
\end{figure}

As was demonstrated in a recent work on predicting solid-state synthesis routes \cite{Aykol2021a}, a candidate set of several recommended reactions can be generated by calculating the Pareto frontier along the reaction distribution shown in Figure \ref{fig:y2mn2o7_rxns}. This is shown in Figure \ref{S-fig:pareto}, and the corresponding reactions are shown in Table \ref{S-tab:pareto}. In particular, the reaction \ce{O2 + 4\ Y3BrO4 + 4\ KMn3O6 -> 4\ KBr + 6\ Y2Mn2O7} stands out as a promising alternative reaction with a zero total chemical potential distance and moderately negative reaction energy. However, it requires \ce{KMn3O6} as a precursor, which to our knowledge, has not been previously experimentally synthesized.

\begin{table}[ht!]
\begin{tabular}{ccc}
\toprule
Reaction &  $\Delta \Phi_\mathrm{rxn}$ (eV/at.) & $\sum \Delta \mu_{\mathrm{min}}$ (eV/at.) \\
\midrule
\ce{4\ YCl3 + 4\ Na3MnO4 -> O2 + 12\ NaCl + 2\ Y2Mn2O7} &  -0.444 &     0.219 \\
\ce{O2 + 4\ YOCl + 4\ NaMnO2 -> 4\ NaCl + 2\ Y2Mn2O7} &  -0.303 &     0.055  \\
\ce{O2 + YMnO3 + 0.5\ RbI -> 0.5\ Y2Mn2O7 + 0.5\ RbIO3} &  -0.258 &     0.035  \\
\ce{O2 + 4\ Y3BrO4 + 4\ KMn3O6 -> 4\ KBr + 6\ Y2Mn2O7} &  -0.159 &     0.000  \\
\ce{3.667\ K2MnO4 + 0.5\ Y2O3 -> O2 + 0.667\ K11Mn4O16 + 0.5\ Y2Mn2O7} &   0.016 &     0.000  \\
\ce{0.2667\ Y(IO3)3 + 0.2667\ MnO2 -> O2 + 0.1333\ Y2Mn2O7 + 0.8\ I}  &   0.439 &     0.409 \\
\bottomrule
\end{tabular}
\caption{Predicted reactions which lie on the Pareto frontier in $\Delta \Phi_\mathrm{rxn}$ and $\sum \Delta \mu_{\mathrm{min}}$ \label{S-tab:pareto}}
\end{table}

\FloatBarrier

\bibliography{supp_r}